\documentclass[12pt]{article}
\usepackage{graphicx}

\begin{document}
\baselineskip 40pt

\vskip 2cm

\def\be{\begin{equation}}
\def\ee{\end{equation}}
\begin{center}
{\Large {\bf Effects of counterrotating interaction on driven tunneling dynamics: coherent destruction of tunneling and Bloch-Siegert shift}}

{\large {\rm Zhiguo L\"{u}$^{1,2}$ and Hang Zheng} $^{1}$}

{$^1$ Department of Physics, Shanghai Jiao Tong University, Shanghai, 200240, China\\
$^2$Institute of Physics and Department of Electrophysics, National Chiao Tung University, Hsinchu 30010, Taiwan}
\date{\today}
\end{center}
\begin{abstract}
We investigate the dynamics of a driven two-level system (classical Rabi model) using the counter-rotating-hybridized rotating wave method (CHRW), which is a simple method based on a unitary transformation with a parameter $\xi$. This approach is beyond the traditional rotating-wave approximation (Rabi-RWA) and more importantly, remains the RWA form with a renormalized tunneling strength and a modified driving strength. The reformulated rotating wave method not only possesses the same mathematical simplicity as the Rabi-RWA but also allows us to explore the effects of counter-rotating (CR) components. We focus on the properties of off-resonance cases for which the Rabi-RWA method breaks down. After comparing the results of different RWA schemes and those of the numerically exact method in a wide range of parameter regime, we show that the CHRW method gives the accurate driven dynamics which is in good agreement with the numerical method. Moreover, the other RWA methods appear as various limiting cases of the CHRW method.  The CHRW method reveals the effects of the CR terms clearly by means of coherent destruction of tunneling and Bloch-Siegert shift. Our main results are as follows: (i) the dynamics of the coherent destruction of tunneling is explicitly given and its dependence on $\Delta$ is clarified, which is quantitatively in good agreement with the exact results; (ii) the CR modulated Rabi frequency and the Bloch-Siegert shift are analytically calculated, which is the same as the exact results up to fourth order; (iii) the validity of parameter regions of different RWA methods are given and the comparison of dynamics of these methods are shown. Since the CHRW approach is mathematically simple as well as tractable and physically clear, it may be extended to some complicated problems where it is difficult to do a numerical study.


\end{abstract}



\pagebreak

\baselineskip 20pt

\section{INTRODUCTION}
\label{intro}
Two-level quantum system (TLS) provides a ideal testing ground for exploring nonclassical phenomena and understanding the nature of quantum physics \cite{rmp,book}. Despite its simplicity, it is an important candidate for modeling diverse situations in nearly every field of physics. The primary importance of a TLS in the fast developing area of quantum information processing is in its controlled manipulation as the elementary building block \cite{Oliver}. Moreover, a series of steps have been taken in this way by time-dependent driving fields. The widely used model here described by the classical Rabi model in the tunneling or localized representation is
\begin{eqnarray}\label{rabi}
    H(t)&=&-\frac{\Delta}{2}\sigma_x - \frac{\epsilon(t)}{2}\sigma_z,  \nonumber  \\
        &=&-\frac{\Delta}{2}\sigma_x - \frac{A}{4}(\sigma_{+} e^{-i \omega t} + \sigma_{-} e^{i \omega t}) - \frac{A}{4}(\sigma_{-}e^{-i \omega t} + \sigma_{+} e^{i \omega t}),
\end{eqnarray}
where $\Delta$ is a time-independent tunneling strength and $\epsilon(t)=A\cos(\omega t)$ is a driving force with amplitude $A>0$ and frequency $\omega$, $\sigma_x$, $\sigma_y$ and $\sigma_z$ are the usual Pauli matrices, $\sigma_{\pm}=(\sigma_{z}\pm i\sigma_{y})/2$.  Since $|s_{1,2}\rangle $ are the eigenstates of $\sigma_x$, then $\sigma_{+} |s_{1}\rangle= |s_{2}\rangle$ and $\sigma_{-} |s_{2}\rangle= |s_{1}\rangle$, where $
|s_{1}\rangle ={\frac{1}{\sqrt{2}}}\left(
\begin{array}{c}
1 \\
1
\end{array}
\right) $ and $|s_{2}\rangle ={\frac{1}{\sqrt{2}}}\left(
\begin{array}{c}
1 \\
-1
\end{array}
\right) $. Thus,
the last two terms in the Hamiltonian Eq.(\ref{rabi}) represent the CR coupling. Subjecting the Hamiltonian to a rotation about the `y' axis, we get a new representation $\exp{(i \pi \sigma_y/4) }H(t) \exp{(-i \pi \sigma_y/4) }= -\frac{\Delta}{2}\sigma_z + \frac{\epsilon(t)}{2}\sigma_x$, which is widely used in quantum optics and nuclear magnetic resonance. Throughout this paper we set $\hbar=1$.

If the CR terms are dropped (in other words, the Rabi-RWA method is applied), Eq.(\ref{rabi}) can be solved exactly. This raises the question on the validity of the results in the strong driving strength and off-resonance regimes, where the breakdown in the RWA occurs \cite{Shirley}. Recently, it has been found that the effects of CR terms are significant in different interesting topics, such as quantum Zeno effect \cite{Zeno}, entanglement evolution \cite{Ficek}, and so on. The main purpose of this paper is to demonstrate the significant role of the CR coupling on the time evolution and physical properties of the driven TLS. We provide a simple analytical method beyond the Rabi-RWA to extract important dynamical features in a wide range of parameter regime. Meanwhile, we also present some methods to find the transition probability for Eq. (\ref{rabi}) without the RWA. After we compare the results obtained by our method with those of the other RWA method and the numerical method, we clearly show the validity of the different methods and prove that the CHRW is more efficient and accurate than the other RWA methods.

The studies of driven TLSs have quite a rich history, and wide application for both experimental and theoretical investigations \cite{Shirley, Hijii,Yuri,kaya,kaya2,Greenberg1,Greenberg2,Son09,Ho09}. Recently, great progress has been made experimentally using superconducting devices based on Josephson tunnel junctions\cite{Tsai,Il,runder,clake,Hof-wang,sun}, optically and electrically controlled single spins in quantum dots\cite{Berezovsky,Koppens,Nowack,xuSC,xuNP}, individual charge in quantum dots\cite{hayashi,Gorman}, and nitrogen vacancy center in diamond\cite{Jelezko,Fuchs,Buckley}, to implement the study of the controllable coherent dynamics of qubits. On the other hand, even for the most simplest driving force with a periodic oscillating field $A \cos(\omega t)$, it is a difficult task to present an analytical and exact solution for this model. Although the model is simple, there appears a wide variety of interesting dynamical features\cite{GH}, like Rabi oscillations, the invalidity of the RWA\cite{Ashhab}, Bloch-Siegert shifts\cite{Shirley,BS-shift,BS-Exp}, non-linear phenomena due to level crossings induced quantum interference, coherent destruction of tunneling (CDT) and the possibility of chaos\cite{CDT1,CDT2,CDT3,CDT4}. In order to discover the driven tunneling physics analytically, a number of approximate methods, such as Rabi-RWA, the RWA in a rotating-frame (RWA-RF) of the Ref.\cite{Ashhab,Hausinger}, have been developed, even though the dynamics of the Rabi model can be solved exactly by numerical methods.

The dynamics of the TLS with far off-resonance and strong driving strength conditions is interesting but difficult to study due to its complexities\cite{Hausinger}. In these approximate methods, the traditional Rabi-RWA ( the CR term $\frac{A}{4}(\sigma_{-}e^{-i \omega t} + \sigma_{+} e^{i \omega t}) $ is neglected ) is one of the most widely used approximations in its study of the resonant interaction between a TLS and a coherent electromagnetic field. This is because it has simple mathematics and possesses a clear physical picture. At the same time it is well known that the Rabi-RWA as a valid method is applicable only to the case of near resonance ($\omega \sim \Delta$) and weak driving strength (or small amplitude $A$). In the driven tunneling problem the former constraint (near resonance) is removed to a certain extent but the latter one (small A) still persists. Though much investigation has been done for more than half a century, obtaining an analytical solution has proved to be an extremely difficult\cite{XH}. Thus some important approximate methods have been developed to study this problem in certain valid parameter regions, such as the RWA-RF method in the Ref\cite{Ashhab}. The RWA-RF method works well in the strong driving field case and gives the resonance properties, but fails in the small oscillation frequency case ($\omega \leq \Delta$) and might be invalid to describe zero static bias case. In order to treat the driven dynamics better, we propose the counter-rotating-hybridized rotating wave method. It possesses the same simple mathematical structure of Rabi-RWA and more importantly, takes into account the effects of CR wave terms. Furthermore, the CHRW method has improved the RWA method in great many cases, and can give the accurate dynamics in a wide range of driving parameter. The obtained results are in good agreement with the exact ones, even in the strong driving strength case, for example the CDT phenomena which is of great interest in the physics of driven quantum tunneling. For a certain condition of the driving parameter, the tunneling is much reduced and even frozen which is called CDT. This results from the effects of CR couplings and happens in some strong driving strength cases. We will show that it is clearly obtained by the CHRW method. Besides, we obtain analytically the exact Rabi frequency and accurate Bloch-Siegert shift up to fourth order by the CHRW method, which manifests the effects of the CR coupling and also strongly support the reasonability and feasibility of this method.

As the Hamiltonian Eq.(1) can be numerically solved easily and quickly, why do we pursue an approximate analytical solution?  The purpose of this work may be: 1) to demonstrate the physics more clearly, such as effects of the CR coupling; 2) to test the accuracy of the analytical solution for extending it to more complicated models where it is difficult to obtain a numerically exact solution; 3) to discuss the validity of the different RWA schemes and show how previous results appear in the various limits of the CHRW method. Hence, we have the following criterion for the validity of an approximate analytical solution: first, it should be as simple as possible, especially mathematically, so that it can be easily extended to more complicated situations where we cannot do a numerical study; second, the main physics should be considered, at least for the interesting and concerned range of the parameters, and it should be as accurate as possible compared with the numerically exact result.

The structure of this article is as follows: we introduce the equations of motion for the Rabi model in Sec. \ref{sec.model}, and then discuss briefly the necessary statement of the numerically exact method to tackle the spin dynamics with an arbitrary condition. In Sec. \ref{sec.CHRWM}, we develop a simple and efficient method to analytically and quantitatively solve the monochromatically driven dynamics. In Sec. \ref{sec.result}, we give the analysis of the dynamics within a wide range of parameter including resonance, near and far off-resonance. Moreover, we demonstrate the effects of CR wave terms on the  dynamics, such as the CDT and the Bloch-Siegert shift. Finally, we sketch a complete diagram of the parameter regions to show the validity of the different approximations, before we present the conclusion.

%

\section{THE NUMERICAL SOLUTION}\label{sec.model}

We give the driven dynamics of the TLS in an arbitrary driving field by a numerical method. The Schr\"{o}dinger equation is
\begin{eqnarray} \label{Sch-eq}
  i\frac{d }{dt}\Psi(t) &=& H(t) \Psi(t).
\end{eqnarray}
The wave function is assumed to be,
\begin{eqnarray*}
  |\Psi\rangle = a(t) \left(
\begin{array}{c}
1 \\
0
\end{array}
\right) + b(t)
\left(
\begin{array}{c}
0 \\
1
\end{array}
\right),
\end{eqnarray*}
where $|a|^2+|b|^2=1$. Substituting the above wave function into Eq.\ref{Sch-eq}, we get the Schr\"odinger equation for the TLS,
\begin{eqnarray}
  i\frac{d a}{dt} &=&  -\frac{\Delta}{2} b - \frac{\epsilon(t)}{2} a,\\
  i\frac{d b}{dt} &=&  -\frac{\Delta}{2} a + \frac{\epsilon(t)}{2} b.
\end{eqnarray}

We define $X=a^{\ast}b+ab^{\ast}$,  $Y=i(a^{\ast}b-ab^{\ast})$, and $Z=a^{\ast}a-bb^{\ast}$.
Using them in Eq.(3) and (4), we write down the following set of equations of motion:
\begin{eqnarray}
  \frac{d X}{dt} &=& -\epsilon(t) Y,   \label{exact-eq1} \\
  \frac{d Y}{dt} &=&  \epsilon(t)X -\Delta Z,   \label{exact-eq2} \\
  \frac{d Z}{dt} &=&  \Delta Y.  \label{exact-eq3}
\end{eqnarray}
By these equations we can give $|a|^2$ or $|b|^2$ in the Schr\"odinger picture.  When $\epsilon(t)$ is dependent on time, it is difficult to give the exact analytical solution in terms of the known special functions due to its complexity\cite{GH,XH}. Luckily, the group of differential equations can be solved by a numerical method for any given initial condition. Thus, in order to obtain the driven dynamics of the TLS, we resort to the fourth order Runge-Kutta method to give the solutions of Eqs.(\ref{exact-eq1},\ref{exact-eq2},\ref{exact-eq3}). At the same time we give the comparison with the results of the other methods like the Rabi-RWA, CHRW, and RWA-RF methods in Sec. \ref{sec.result}. In principle, we can give the dynamics and physics of the model in the whole parameter regime. In this article we consider the situation of the driven TLS with zero static bias and $\omega \neq 0$.



\section{COUNTER-ROTATING HYBRIDIZED ROTATING WAVE METHOD} \label{sec.CHRWM}


In this section, we apply a unitary transformation to calculate the driven tunneling dynamics\cite{Zeno,ZL}. We propose that the generator of the unitary transformation is $S=-i\frac{A}{2\omega}\xi\sin(\omega t)\sigma_z $, a time dependent operator. The parameter $\xi$ is introduced in $S$ and will be determined later on. Thus we use the time-dependent Schr\"odinger equation Eq. \ref{Sch-eq} to solve the dynamics.
After the unitary transformation, we obtain readily the interaction picture formulas with $\Psi'(t)=\exp(S)\Psi(t)$ and $i\frac{d }{dt}\Psi'(t) = H'(t) \Psi'(t) $, where
\begin{eqnarray}\label{H'}
  H' &=& -\frac{\Delta}{2}\left[\sigma_x \cos\left( \frac{A}{\omega}\xi\sin(\omega t)\right) + \sigma_y \sin\left( \frac{A}{\omega}\xi\sin(\omega t)\right)\right] \\ \nonumber
   && -\frac{A}{2}\left[1-\xi\right]\cos(\omega t) \sigma_z .
\end{eqnarray}
After making use of the relation given in Ref.\cite{Ashhab}
\begin{eqnarray}
  \exp(iz \sin\alpha) &=& \sum^{\infty}_{n=-\infty}J_n(z) e^{in \alpha},
\end{eqnarray}
where $J_n(z)$ are Bessel functions of the first kind, we rewrite Eq.(\ref{H'}) as
\begin{eqnarray} \label{Hj'}
  H' &=&  -\frac{\Delta}{2}\sum^{\infty}_{n=-\infty} J_n\left(\frac{A}{\omega}\xi\right) \left[ \cos(n \omega t)\sigma_x + \sin(n \omega t)\sigma_y \right] \\ \nonumber
  && -\frac{A}{2}\left[1-\xi\right]\cos(\omega t) \sigma_z .
\end{eqnarray}
We divide the Hamiltonian into three parts $H'=H'_0+H'_1+H'_2$.

\begin{eqnarray}
  H_0'&=& -\frac{\Delta}{2}J_0\left(\frac{A}{\omega}\xi\right)\sigma_x, \\
  H_1'&=& - \frac{\Delta}{2}2J_1\left(\frac{A}{\omega}\xi\right)\sin(\omega t)\sigma_y-\frac{A}{2}(1-\xi)\cos(\omega t)\sigma_z,
\end{eqnarray}
and $H_2'=H'-H_0'-H_1'$ includes all higher order harmonic terms, such as $\cos(2\omega t), \sin(2\omega t)$.  Then, in order to make the CR wave terms in $H_1'$ disappear, we should choose a proper parameter $\xi$, which satisfies
\begin{eqnarray} \label{xi}
  0 &=& \frac{A}{2}(1-\xi)-\frac{\Delta}{2}2J_1\left(\frac{A}{\omega}\xi\right).
\end{eqnarray}
Notice that from Eq.(\ref{H'})  to Eq.(\ref{xi}) no approximation is involved.  In the following treatment, we neglect the higher order harmonic terms of $H_2'$ ($n \omega$, $n=2,3,4...$) and thus, we obtain our reformulated rotating-wave Hamiltonian which is called the counter-rotating-hybridized rotating wave method

\begin{eqnarray}\label{CHRW}
 H_{\mathrm{CHRW}} &=& H_0'+H_1'=-\frac{\tilde{\Delta}}{2}\sigma_x -  \frac{\tilde{A}}{4}(\sigma_{+}\exp(-i \omega t) + \sigma_{-} \exp(i \omega t)),
\end{eqnarray}
where $\tilde{\Delta}=\Delta J_0\left(\frac{A}{\omega}\xi\right)$ is the renormalized tunneling strength and $\tilde{A}=2A (1-\xi)=4 \Delta J_1\left(\frac{A}{\omega}\xi\right)$ is the renormalized amplitude of the driving field.  It is obvious to see that the effects of CR terms have been taken into account in our treatment which leads to the renormalization of the physical properties. An interesting point about the CHRW method is that the CHRW Hamiltonian has the same mathematical formulation as the Rabi-RWA one except for the renormalized physical quantities. Therefore, it is easy to obtain the CHRW dynamics as well as the Rabi-RWA one.

In the following, we calculate the occupation probability $P_{\mathrm {up}}(t)$ \cite{Ashhab} in both the CHRW and Rabi-RWA Hamiltonians. $P_{\mathrm {up}}(t)$ denotes the probability that a system under an initial down state of $\sigma_z$ ($P_{\mathrm {up}}(0)=0$), is in the up state of $\sigma_z$ at time $t$. As it is readily obtained by many approaches, we use the Schr\"odinger equation method to get it. Besides, because the generator $S$ is a function of $\sin(\omega t)$, the initial condition of the wave function is invariant after the unitary transformation, namely, $\Psi'(0)=\Psi(0)$. The wave function in the basis $|s_{1,2}\rangle$ is
\begin{equation}\label{wave}
    \Psi'(t)=c_1(t) |s_{1}\rangle +c_2(t) |s_{2}\rangle.
\end{equation}
Using the Schr\"odinger equation, we have
\begin{eqnarray}
  i\frac{d c_{1}}{dt} &=&  -\frac{\tilde{\Delta}}{2} c_{1} -\frac{\tilde{A}}{4} e^{ i\omega t} c_{2},  \\
  i\frac{d c_{2}}{dt} &=&   \frac{\tilde{\Delta}}{2} c_{2} -\frac{\tilde{A}}{4} e^{-i\omega t} c_{1}.
\end{eqnarray}
On setting  $\widetilde{c_{1}}=c_{1}e^{-i\tilde{\Delta} t/2}$, $\widetilde{c_{2}}=c_{2}e^{i\tilde{\Delta} t/2}$, we immediately obtain
\begin{eqnarray}
  i\frac{d \widetilde{c_{1}}}{dt} &=&  -\frac{\tilde{A}}{4} e^{ i(\omega-\tilde{\Delta}) t} \widetilde{c_{2}}, \label{tidec1} \\
  i\frac{d \widetilde{c_{2}}}{dt} &=&  -\frac{\tilde{A}}{4} e^{-i(\omega-\tilde{\Delta}) t} \widetilde{c_{1}}. \label{tidec2}
\end{eqnarray}
These equations give us a complete solution of the problem. All the physically relevant quantities can be obtained from Eq.(\ref{tidec1}) and Eq.(\ref{tidec2}). We find that $\widetilde{c_{1}}$ satisfies a linear second-order differential equation with constant coefficients\cite{GH}. For any initial conditions,  for example, $c_1(0)=c_2(0)=\frac{1}{\sqrt{2}}$, i.e.$\langle\sigma_z(0)\rangle=1$, we can readily solve the equations Eq.(\ref{tidec1}) and (\ref{tidec2}) as given in Ref. \cite{scu},
\begin{eqnarray}
  c_{1}(t) &=&   e^{ i\frac{\omega t}{2}} \left\{\frac{1}{\sqrt{2}}\cos\left(\frac{\tilde{\Omega}_R t}{2}\right)+i \frac{\frac{\tilde{A}}{2}-\tilde{\delta}}{\sqrt{2}\tilde{\Omega}_R} \sin\left(\frac{\tilde{\Omega}_R t}{2}\right)  \right\},   \\
  c_{2}(t) &=&  e^{ -i\frac{\omega t}{2}} \left\{\frac{1}{\sqrt{2}}\cos\left(\frac{\tilde{\Omega}_R t}{2}\right)+i \frac{\frac{\tilde{A}}{2}+\tilde{\delta}}{\sqrt{2}\tilde{\Omega}_R} \sin\left(\frac{\tilde{\Omega}_R t}{2}\right)  \right\},
\end{eqnarray}
where $\tilde{\delta}=\omega-\tilde{\Delta}$ is the renormalized detuning parameter and $\tilde{\Omega}_R=\sqrt{\tilde{\delta}^2+\tilde{A}^2/4}$ is the modulated Rabi frequency of the CHRW method. Thus the population of the state $\left(
\begin{array}{c}
0 \\
1
\end{array}
\right)$ in the $\sigma_z$ basis is
\begin{eqnarray}\label{pt-CHRW}
P_{\mathrm {up}}^{\mathrm{CHRW}}(t) &=& \left( \frac{\tilde{A}}{2\tilde{\Omega}_R}\right)^2\sin^2\left(\frac{\tilde{\Omega}_R t}{2}\right)\sin^2\left(\frac{\omega t}{2}\right)+ \\ \nonumber
&&\left[ \cos\left(\frac{\tilde{\Omega}_R t}{2}\right)\sin\left(\frac{\omega t}{2}\right)-\frac{\tilde{\delta}}{\tilde{\Omega}_R} \sin\left(\frac{\tilde{\Omega}_R t}{2}\right)\cos\left(\frac{\omega t}{2}\right)\right]^2.
\end{eqnarray}
The renormalized Rabi frequency $\tilde{\Omega}_R$ has taken into account the effects of CR terms on frequency shifts and will give the Bloch-Siegert shift in a simple way (we will show the calculation in Sec. \ref{sec.result}). Besides, We clarify the relation between the CHRW approach and the other RWA methods in Sec. \ref{sec.sum}.

The Hamiltonian of the Rabi-RWA method reads,
\begin{equation}\label{rwa}
    H_{\mathrm{Rabi}}=-\frac{\Delta}{2}\sigma_x - \frac{A}{4}(\sigma_{+}\exp(-i \omega t) + \sigma_{-} \exp(i \omega t)).
\end{equation}
Since we have the CHRW solution, we can immediately obtain the Rabi-RWA result,
\begin{eqnarray}\label{pt-rwa}
  P_{\mathrm {up}}^{\mathrm{Rabi}}(t) &=& \left( \frac{A}{2\Omega_R}\right)^2\sin^2\left(\frac{\Omega_Rt}{2}\right)\sin^2\left(\frac{\omega t}{2}\right)+ \\\nonumber
  &&\left[ \cos\left(\frac{\Omega_Rt}{2}\right)\sin(\frac{\omega t}{2})-\frac{\delta}{\Omega_R} \sin\left(\frac{\Omega_Rt}{2}\right)\cos\left(\frac{\omega t}{2}\right)\right]^2.
\end{eqnarray}
where $\delta=\omega-\Delta$ is the detuning parameter and $\Omega_R=\sqrt{\delta^2+A^2/4}$ denotes the Rabi freqency of the Rabi-RWA method.

Finally, we show the result of the RWA-RF method\cite{Hausinger} is a limit case of the CHRW method. From Eq.(\ref{xi}), we get $\xi=1$ in the limit of $A \rightarrow \infty$, and simultaneously obtain the Hamiltonian of the rotating frame\cite{Ashhab} by Eq.(\ref{Hj'}).  Notice that there still exist counter-rotating couplings in $\sigma_y$ term though the $\sigma_z$ term has been eliminated in this case. For the resonance condition $|n\omega|=0$ to hold, only one value of $n$ is kept. Therefore, the effective RWA-RF Hamiltonian is written as
\begin{equation}\label{rwa-rf}
    H_{\mathrm{RWA-RF}}=- J_0\left(\frac{A}{\omega}\right) \frac{\Delta }{2}\sigma_x .
\end{equation}
Thus, the probability $P_{\mathrm {up}}(t)$ of the RWA-RF approach in Ref.\cite{Ashhab} is obtained,
\be \label{rrwa}
P_{\mathrm {up}}^{\mathrm {RWA-RF}}(t)=\sin^{2}\left\{J_0\left(\frac{A}{\omega}\right)\frac{\Delta t}{2}\right\},
\ee
whose amplitude is always one for any driving parameter. This means that the $P_{\mathrm {up}}(t)$ always exhibits a full periodic oscillation between the up and down states of $\sigma_z$ except the CDT case ($P_{\mathrm {up}}^{\mathrm {RWA-RF}}(t)\equiv 0$ due to $J_0\left(\frac{A}{\omega}\right)=0$). Therefore, the result Eq. (\ref{rrwa}) of the RWA-RF is distinguished from that of the CHRW (Eq.(\ref{pt-CHRW})). This treatment simply corresponds to the case $\xi=1$ of the CHRW method, which is only valid in the limit of really strong driving strength $A \gg \omega, \Delta$. Furthermore the derivation of the modified tunneling $\Delta J_0(A\xi/\omega)$ and rescaled driving strength $\tilde{A}$ is extremely significant in Eq.(\ref{CHRW}). By quantitative comparisons with numerical results it is shown that for some parameters the CHRW method gives a significantly better description than the previous treatments. The discussion of the CR effects and comparisons of these RWA dynamics with the numerically exact ones are shown in the following section.

In summary, the present work provides a new method, a novel kind of renormalized scheme based on a unitary transformation, to study driven quantum dynamics. After the unitary transformation, we obtained the renormalized RW form with mathematical simplicity and physically clear picture. Simply speaking, one can compare the physical quantities of the original model with those of the transformed model,
\begin{itemize}
  \item A $\longrightarrow$ \~{A}=2A(1-$\xi$);
  \item {$\Delta$} $\longrightarrow $ $\widetilde {\Delta}=\Delta J_0\left(\frac{A}{\omega}\xi\right)$;
  \item $\Omega_R$ $\longrightarrow$ $\widetilde{\Omega}_R$.
\end{itemize}
The renormalized quantities in the transformed Hamiltonian Eq. (\ref{CHRW}) results explicitly from the effects of CR interactions. Physically, the renormalized effects can be detected from the Bloch-Siegert shift. More obviously and straightforwardly, the time evolution of the driven TLS is sensitive to the renormalized quantities. In the renormalized rotating-wave framework, we would demonstrate it by the driven TLS dynamics of different parameter regions in the next section.

\section{RESULTS AND DISCUSSION}
\label{sec.result}
We systematically discuss the dynamics of different parameter ranges: resonance, near resonance, and far off-resonance. Increasing the driving strength from the weak coupling regime to the strong coupling case we will see a rich distinct dynamics. At the same time we compare all the results of the CHRW approach with those of the other methods, namely, the Rabi-RWA method, the numerically exact method, the RWA-RF method. More importantly, we give significant discussions about the interesting phenomena CDT and the Bloch-Siegert shift which clearly emerges in the strong coupling or large detuning case. Finally we give an overview of the parameter regime in which the different approaches are valid. We discuss how previous RWA results appear in the various limits of the CHRW method.

\subsection{Resonance and near resonance}

Let us take a look at the resonance ($\omega=\Delta$) and near resonance ($\omega\sim\Delta$) dynamics from the weak driving strength to the strong driving strength. In Fig. 1, we show the occupation probability $P_{\mathrm{up}}(t)$ at resonance with $A/\omega \geq 1$. For comparison, we also give the results of the other different approaches. It is easy to check that for a very weak driving strength the results of all methods are nearly the same. However, for $A/\omega>2$, the Rabi-RWA method breaks down, while our method still works quite well and $P^{\mathrm{CHRW}}_{\mathrm{up}}$ is in quantitatively good agreement with numerically exact results. It is obvious to see that, for the RWA-RF results $P^{\mathrm{RWA-RF}}_{\mathrm{up}}$, the deviation from the exact results becomes much larger with the increase of the driving strength, which is clearly seen in Fig. 1 from $A/\omega = 1$ to $A/\omega = 2.5$. In contrast, the CHRW results agree well with the exact results even for $A/\omega=2.5$.

In the near resonance and moderate driving strength cases ($\omega=1.2 \Delta$ and $A/\omega \geq 1$, see Fig.2), the Rabi-RWA dynamics exhibits distinct differences from the exact one. However, our results are in good agreement with the  numerically exact results. Besides, one can observe the coherent fast oscillation from Fig. 2 meanwhile the modulated amplitudes clearly exhibits a slow oscillation (see Fig. 2b). Physically, the driven tunneling dynamics has two intrinsic frequencies: one is the Rabi frequency and the other is the driven frequency. The results of the CHRW method illustrate this character and agree very well with the exact results. However, the RWA-RF method gives only a coherent oscillating character with a single oscillatory frequency and shows that the amplitude of $P^{\mathrm{RWA-RF}}_{\mathrm{up}}(t)$ is always one. It is because that the only term fulfilling the resonance condition $n\omega=0$ remains while the other rotating components are dropped (See Eq.\ref{rrwa}). Thus, for near resonance case $\omega > \Delta$, the RWA-RF dynamics is totally different from the exact one as the Rabi-RWA does.

We show the dynamics of near resonance $\omega<\Delta$ for different approaches in Fig.3. When $1<\Delta/\omega<1.2$, the Rabi-RWA works for $2\geq A/\omega$ with small errors of both amplitude and frequency. However, it breaks down in the strong coupling case. From the calculated results, one can see that the Rabi-RWA treats the detuning case ($\omega<\Delta$, Fig. 3a) better than  the other detuning case ($\omega>\Delta$, Fig. 2a). However, in both the near resonance cases the RWA-RF gives the perfect oscillatory behaviors with a longer period which is larger than the exact result (Fig. 3b).  In comparison with the exact results, the CHRW performs quite well at resonance and near resonance even in moderately strong driving strength regime, for example, $A/\omega=2.5$.

From all the above figures, it is clear to see that the CHRW obtains correctly the novel dynamical characters of on-resonance and near-resonance. By comparison with the numerical results it is shown that the CHRW treatment gives a significantly better description than the previous treatments.
Physically, it treats the CR and rotating wave terms on equal footing. Moreover, the CR coupling results in the renormalization of physical quantities in the CHRW Hamiltonian (see Eq. (\ref{CHRW})). For example, in the near resonance case of $A/\omega=2$ and $\omega/\Delta=1.2$ ( Fig. 2b ), we get $\xi=0.5894$ by solving Eq.\ref{xi}. Thus we obtain the renormalized physical quantities $ \widetilde{A}= 1.9711 \Delta$, $\widetilde{\Delta}= 0.6817 \Delta$. One can see that the time evolution of the CHRW method is quantitatively in good agreement with the numerically exact result, but the Rabi-RWA and RWA-RF results show large deviation from the exact result. Due to the renormalization, the dynamics $P_{\mathrm{up}}(t)$ of the TLS (Eq. (\ref{pt-CHRW})) not only recovers the usual form of the Rabi-RWA method (See Eq. (\ref{pt-rwa}) ) but also give the correct driven tunneling dynamics. Thus it is a simply tractable method. More importantly, it allows us to study the influence of the CR terms on the dynamics and the physics in the invalid parameter regime of the Rabi-RWA method, especially the moderately strong driving strength and the far off-resonance regimes.

\subsection{Far off-resonance and CDT}

From weak to moderately strong driving strength regime, the results obtained by the CHRW method in the off-resonance case agree well with the numerically exact ones. The traditional Rabi-RWA method can describe the physics in the weak driving strength and the near resonance cases but is invalid in the far off-resonance case and beyond the weak driving strength regime. In contrast, the CHRW approach works well in all the cases. We show the dynamics $P(t)$ for $\Delta/\omega=0.2$ with different driving strengths in Fig. 4(a-d) ((a)$A/\omega=1$, (b)$A/\omega=2$, (c)$A/\omega=2.5$, and (d)$A/\omega=6$, respectively), and show the other detuning case $\Delta/\omega=6$ in Fig. 4(e) ($A/\omega=2$) and Fig. 4(f) ($A/\omega=6$). In the ultra-strong driving strength regime with far red detuning $\Delta/\omega \gg 1$ (see Fig. 4(f)), the CHRW results are qualitatively in agreement with the numerically exact results. When $\Delta/\omega <$ or $\sim 1$, even in the strong driving strength regime, the CHRW results are consistent with the numerically exact ones (See Fig.4(d), $A/\omega=6$). But a large discrepancy emerges between the results of the RWA-RF method and those of the exact method for a weak driving strength and $\Delta>\omega$, which is most clearly seen in Fig. 4(e) and Fig. 4(f). It is obvious to see that the sinusoidally oscillating function in Eq. (\ref{rrwa}) agrees with the main oscillation of the exact result for the strong driving strength regime and $\Delta = 0.2 \omega$ (See Fig.4(a-d)). However, the exact numerical dynamics exhibits a distinct characteristic, a remarkably oscillatory structure in the short-time dynamics. From the insets of Fig. 4, the short time evolution of the exact results displays some fine structures consisting of fast oscillations, while that of the CHRW method shows small wiggly oscillations, and that of the RWA-RF method show no oscillation but the monotonic smooth increase. Physically, the fine structure of the dynamics, especially in the strong driving strength case, comes from the effects of the harmonic terms ($n\omega$, $n=1,2,3,...$) which were included by $H'_1+H'_2$. Since the CHRW method has taken into account the effects of $H'_0+H'_1$, it gives rise to some characteristics of the fine structure as well as the main oscillation of the envelope. We will discuss the validity of the different methods at the end of the next subsection.

The CDT, an intriguing phenomenon of suppression of coherent tunneling in the TLS dynamics, has been found\cite{CDT1}, which comes from the effects of CR couplings on the dynamics. We demonstrate the well-known CDT when $\omega \gg \Delta$ and $A/\omega=2.4048, 5.03$ (zero points of the zero-order Bessel function). In Fig. 5, we set $A/\omega=2.4048$ and show the CHRW dynamics for $\omega/\Delta=2$ (Fig. 5(a)), $\omega/\Delta=6$ (Fig. 5(b)), $\omega/\Delta=10$ (Fig. 5(c)), $\omega/\Delta=20$ (Fig. 5(d)), respectively. The results of the other methods are also shown for comparison in these figures. The main results are summarized in Table \ref{tab:hresult}. Since $J_0\left(\frac{A}{\omega}=2.4048 \right)\equiv 0$ in Eq.\ref{rrwa}, the RWA-RF result $P_{\mathrm {up}}^{\mathrm {RWA-RF}}=0$ is independent of the value of $\omega/\Delta$. It is important to notice that when $\omega \sim \Delta$ (See Fig. 5(a)), there is no CDT but a long-periodic oscillation superposed with small-amplitude fast oscillations obtained by the CHRW methods, which is in good agreement with the result of the numerically exact method. These fast oscillatory characteristics result from the multi-$\omega$ terms of Eq.(\ref{Hj'}) ($n \omega, n=1,2,3,...$). The CHRW results indicate that, as $\omega/\Delta$ increases from $2$ to $20$, the driving induced suppression of tunneling emerges clear only for $\omega/\Delta \geq 6$ (See Fig. 5(a)-(d)). There is still a small-amplitude fast oscillation for $\omega/\Delta = 6$ in the short-time evolution and the long time evolution $\Delta t \gg 10$ exhibits a very long periodic oscillation (It is not shown in Fig. 5(b)).  Until $\omega/\Delta \geq 10$ the amplitude is extremely small (See Fig. 5(c) and Fig. 5(d)). For $A/\omega=2.4048$ and $\omega/\Delta=20$, we get $\xi=0.9780$ by self-consistently solving Eq. (\ref{xi}). Thus, we obtain the renormalized physical quantities $ \widetilde{A}= 2.1205 \Delta$, $\widetilde{\Delta}= 0.0278 \Delta$. It is obvious to see the renormalized effects on the dynamics $P(t)$ by the CHRW method, i.e., the large suppression of the amplitudes in Fig. 5(d), in comparison with the results of the Rabi-RWA method. It means that it is necessary to take into account the CR interactions in order to illustrate correctly driven tunneling dynamics, especially, in strong coupling regime or far off-resonance case. On the other hand, there emerges a periodic oscillation with a small amplitudes (In Fig.5(c), the amplitude is about $0.012$; in Fig. 5(d), the amplitude is about $0.003$.), which agrees well with the numerically exact result, while the RWA-RF result is $P(t)=0$.  Therefore, only when $\omega/\Delta \gg 1$ and for $A/\omega$ satisfying the condition $J_0\left(\frac{A}{\omega}\xi\right)\simeq 0$, the amplitude of the oscillation tends to zero. It is worth noticing that only for this condition does the TLS remain its initial state forever, which indicates that a dynamical broken symmetry happens. However, the Rabi-RWA results are totally different from the exact ones for $A/\omega=2.4048$.

We show the condition of the CDT in the following. From the literature\cite{Ashhab,CDT1,CDT2,CDT3,CDT4}, people usually believes that the tunneling of the TLS is only determined by the factor $J_0(A /\omega)$, which leads to a reduction of the effective tunneling and even the complete destruction for $A/\omega$ satisfying $J_0(A /\omega)=0$. It is obvious that this condition is independent of $\Delta$. While, analytical results of the CHRW method turn out that the condition of the CDT is $J_0(A \xi /\omega)=0$ which is confirmed by the numerically exact results. Furthermore, if $A/\omega=2.4048$, the CDT predicted by the RWA-RF method is only determined by $\Delta J_0(A /\omega)=0$ in Ref.\cite{GH,Ashhab}, but if $J_0(A \xi /\omega)\neq 0$  there is no CDT. For example, when $\omega \sim \Delta$, a periodic oscillation with a large amplitude appears(See Fig. 5(a)). Moreover, for $\omega \sim 20 \Delta$ an oscillation with a small nonzero amplitude exists. The results of the CHRW method agree well with the numerical results, as is shown in Fig. 5(d). At the same time the results of the RWA-RF method in Fig. 5 show $P_{up}\equiv 0$. For the Rabi-RWA method, because of no $J_0(A \xi/\omega)$ in Eq.(\ref{rwa}), its dynamics exhibits highly oscillatory with large amplitudes.

Finally, let us discuss the difference of the CDT condition between the CHRW treatment and RWA-RF method. For the CHRW method, when $A \gg \omega$ and $\omega\gg\Delta$, $\xi$ tends to $1$ by Eq.(13), and then $A\xi/\omega $ goes to $A/\omega$. Thus the CDT occurs if $A/\omega\xi$ satisfies $J_0(\frac{A}{\omega} \xi)=0$. However, according to the probability $P(t)$ of the RWA-RF approach in Ref.\cite{Ashhab}, as the value of ${A}/{\omega}$ is a zero point of zero-order Bessel function $J_0$, $P_{\mathrm{up}}(t)$ remains zero forever. In other words, for a symmetric TLS, the occurrence of the CDT phenomenon obtained by the RWA-RF method is irrelevant to the value of $\Delta$. From our analysis we know that this condition can be invalid for the lower driving frequency case $\omega/\Delta < 10 $. In Fig. 5(c) and Fig. 5(d), it is seen that only when the driving frequency $\omega$ is much higher than $\Delta$ and $\frac{A}{\omega}\xi$ is a zero point of $J_0$, the CDT occurs. Therefore, the renormalized effects of the CR terms should be properly taken into account in order to describe the physics of driven tunneling dynamics, especially zero bias case\cite{Ashhab}.

\subsection{Bloch-Siegert shift and validity of different RWAs}
In the subsection we calculate the Bloch-Siegert shift. The shift is a well known correction to the RWA, and accounts for the CR field to leading order. The renormalized Rabi frequency is
\begin{eqnarray}
  \tilde{\Omega}_R^2 &=& \left[\omega - \Delta J_0\left(\frac{A\xi}{\omega}\right)\right]^2 + \left[2 \Delta J_{1}\left(\frac{A\xi}{\omega}\right)\right]^2.
\end{eqnarray}
Then, we expand $J_0\left(\frac{A\xi}{\omega}\right)$ and $J_1\left(\frac{A\xi}{\omega}\right)$ up to fourth order in $A$
\begin{eqnarray}
  J_{0}\left(\frac{A\xi}{\omega}\right)&=& 1- \frac{1}{4}\left(\frac{A\xi}{\omega}\right)^2 + \frac{1}{64}\left(\frac{A\xi}{\omega}\right)^4+O(A^6), \\
 2\Delta J_{1}\left(\frac{A\xi}{\omega}\right)&=& \Delta\left [\frac{A\xi}{\omega}-\frac{1}{8}\left(\frac{A\xi}{\omega}\right)^3+ O(A^5)\right],
\end{eqnarray}
and expand $\xi$ up to second order in $A$,
\begin{eqnarray}
 \xi&=&\frac{\omega}{\omega+\Delta}\left(1+\frac{A^2}{8}\frac{\Delta}{(\omega+\Delta)^3}+O(A^4)\right).
\end{eqnarray}
Thus the modulated effective Rabi frequency gives

\begin{eqnarray}\label{Rabifreq}
  \tilde{\Omega}_R^2 &=& \left[\omega-\Delta\right]^2+\frac{A^2\Delta}{2(\omega+\Delta)}- \frac{A^4\Delta}{32(\omega+\Delta)^3},
\end{eqnarray}
which is the same result as those of Refs. \cite{GH} and \cite{BS-shift}.  $\tilde{\Omega}_R$ has taken into account the CR couplings. Morover, Eq. (\ref{Rabifreq}) can be used to calculate the Bloch-Siegert shift of the resonance frequency $\Omega_{\mathrm{res}}$. It is defined as the frequency at which the transition probability $P_{\mathrm {up}}$ averaged is a maximum. This occurs when $\partial \tilde{\Omega}_R^2/ \partial \Delta =0$\cite{BS-shift}. Thus, we obtain the Bloch-Siegert shift $\delta\omega_{\mathrm {BS}}$,
\begin{equation}\label{BS}
    \delta\omega_{\mathrm {BS}}=\Omega_{\mathrm{res}}-\Delta=\frac{1}{16}\frac{A^2}{\Delta}+\frac{(A/4)^4}{4\Delta^3}.
\end{equation}
This is the exact result as given in Ref.\cite{BS-shift}. These results strongly prove that the CHRW method has properly taken into account the effects of CR terms.

It is significant to note that different approaches have different parameter regions of validity, as shown in Fig. 6. We illustrate the valid region of the conventional RWA method (the weak coupling and near resonance region), whose dynamics is most clearly shown before (Rabi physics). Physically, since the Rabi-RWA has not considered the effects of the CR interactions, it can not be used to explore the strong driven tunneling dynamics or far off resonance dynamics, such as the CDT and Bloch-Siegert shift. It is obvious to see that the parameter region of validity for the generalized RWA method, i.e. the CHRW method covers that for the Rabi-RWA method, and the sparse shaded region is much larger and broader than the dense shaded region of the Rabi-RWA method. It works very well when $\omega>\Delta$, even for ultra-strong driving strength case when $\omega \gg \Delta$, and also gives the accurate results for $A/\omega \leq 2$ when $\omega\leq\Delta $. Note that for the red zone parameter the CDT occurs. As mentioned above, the RWA-RF method predicts the occurrence of the CDT is irrelevant to the ratio of $\Delta/\omega$. Further, the RWA-RF method can give the strong driving physics in the zero bias case ($A \gg \omega, A\gg \Delta, \epsilon(t)=A\cos(\omega t)$). By comparison, it is seen that the RWA-RF results exhibit a large deviation from the exact results for all off-resonance cases $\omega < \Delta$.

Beyond the sparse region and its neighborhood, the CHRW results are still in qualitatively good agreement with exact results in respect of overall changes of amplitudes and oscillation frequency. However, there is a slight discrepancy for $\Delta/\omega \geq 2$ and strong driving case (for instance, $A/\omega=3$ and $\Delta/\omega=2$). We believe that the discrepancy indicates the inadequate treatment of the CR terms which gives rise to the loss of a part of the physics in the regime. It is because that the CHRW method does not take into account higher order harmonics (multi-photon transition processes) which have important roles in the case of $\Delta/\omega \gg 1$ and $A/\omega \gg 1$. While in the regime $\Delta/\omega \leq 1$ the effects of these processes are not prominent, thus the CHRW method can give the accurate result that is nearly the same as the exact one. Besides, the renormalized tunneling $\tilde{\Delta}$ in $H'_0$ and the renormalized driving strength $\tilde{A}$ in $H'_1$ have properly taken into account the effects of the driving field. In consequence, our results turn out that it is necessary to treat both $H'_0$ and $H'_1$ on equal footing to obtain accurate effects owing to the CR couplings. In contrast, the RWA-RF method only considers the renormalized tunneling, whereas the Rabi-RWA method neglects the CR terms from the beginning.

\section{SUMMARY}\label{sec.sum}


To summarize, we systematically study the driven tunneling dynamics of a TLS under a periodic driving field using the CHRW method, which is based on a unitary transformation. This method treats the rotating wave terms and the CR ones on equal footing. Thus, we not only gives the weak driving strength results, such as the Rabi physics, but also show the strong driving strength results, such as the CDT phenomena. Physically, all the results are dependent on the renormalized tunneling $\tilde{\Delta}$ and renormalized driving strength $\tilde{A}$. Within the framework of the CHRW method, the interaction between the driving field and the TLS leads to the renormalization of the two parameters. It allows us to calculate the analytical properties of driven tunneling dynamics in the renormalized rotating-wave framework. Meanwhile, from the driven tunneling dynamics, it is seen that the characteristics of the oscillations are very sensitive to $\widetilde A$ and $\widetilde {\Delta}$. In comparison with the other analytical methods, the CHRW method is a simple and feasible treatment since it holds the RWA mathematical form.  More importantly, the CHRW approach can give the dynamics in good agreement with exact results within a wide region of parameter space. Unlike the conventional Rabi-RWA method, this technique is nonperturbative, so that it can be applied to study the driven tunneling physics in a broad region of parameter, especially beyond the weak coupling regime and the near resonance case. In a wide range of values corresponding to $\Delta$, $A$ and $\omega$ parameters, we compare different RWA schemes and the numerically exact method. The results indicate that the CHRW method gives the accurate driven dynamics in the parameter regimes ($\Delta/\omega<1$, $A/\omega$ from small values to much larger than one) and ($\Delta/\omega>1$, $A/\omega \leq 2$ ), and in the neighboring regimes the driven tunneling dynamics though not exact is in good agreement with that of the numerical method. Moreover, the other RWA methods appear as various limiting cases of the CHRW method. By the CHRW method, the CDT induced by the driving field is obtained, which mimics a local quasi-equilibrium and generates a phase with broken symmetry. We demonstrate that the condition of the CDT is dependent on both the value of A and the ratio $\Delta/\omega$. Further comparisons with the numerically exact results and the calculation of the Bloch-Siegert shift are explicitly given, which strongly proves the reliability of the CHRW method. Moreover, our findings allow us to speculate about some general features and significant effects of the CR wave terms behind the phenomenon of driven tunneling dynamics. It might account for the versatile strongly-driven experiments investigated in \cite{Fuchs,BS-Exp}. The method can be applied to more complicated driving problems and dissipative dynamics exposed by strong ac driving\cite{GH}.

We would like to write down a few words about the treatment of neglecting $H'_2$. After the unitary transformation, we obtained the transformed Hamiltonian and then divided it into three parts $H'_0$, $H'_1$ and $H'_2$.  $H'_0$ is the renormalized tunneling with the renormalized factor $J_0(A\xi/\omega)$ including an infinite order of $A$. $H'_1$ contains all single-$\omega$ terms which relates to single-photon assisted transitions. Moreover,$H'_1$ possesses the RWA form after we choose $\xi$ by the self-consistent equation Eq. (13). $H'_2$ involves all multi-$\omega$ terms ($n \omega$, $n=2,3,4...$), which corresponds to the multi-photon assisted transitions. Our calculations of neglecting $H'_2$ work well even for the moderately strong driving strength region. It is justified by our following results, (i) by the CHRW method we obtain the CDT dynamics which agrees well with the exact results; (ii) we calculate the modulated Rabi frequency and the Bloch-Siegert shift. The results to fourth order are the same as the exact results; (iii) as $H'_2$ includes all the multi-$\omega$ terms ($n \omega$, $n=2,3,4...$), its contribution to the dynamics is not prominent except for the ultra-strong driving strength case. This point has been verified by our calculations of the $P(t)$ in comparison with the numerically exact results from weak to moderately strong driving strength.  After we compare the CHRW results with those of the Rabi-RWA method, it turns out that the valid parameter region of the CHRW method is much more broader than that of the Rabi-RWA method which is perturbative in the driving strength [39].  We figure the regions of validity for the different methods in Fig. 6. In this figure, we can see clearly that the CHRW method covers the large areas of the parameter region in which the RWA method breaks down. All these results conclusively support the reasonability of neglecting $H_2'$.

We would like to clarify the relation between the CHRW approach and the other RWA methods. The key point of our treatment is the unitary transformation with the generator $S=-i\frac{A}{2\omega}\xi\sin(\omega t)\sigma_z $, where a parameter $\xi$ is introduced. After the transformation an expansion has been performed according to the order of the harmonics (photon transfer process: $0$ photon, $1$ photon,$...$, $m$ photons). The CHRW treatment does not consider the process of higher order harmonics ($m \geq 2$). If $\xi=0$, in other words, without the transformation, the expansion would give the Rabi-RWA form with neglecting the CR terms. While $\xi=1$, our transformation is the usual polaron-like transformation which gives the rotating-frame form\cite{Oliver,Ashhab}, and if we use the resonance condition, then we can obtain the results of the RWA-RF method. It is important to notice that the CR terms exist in the sine functions (See Eq.(\ref{Hj'})). Our choice in the CHRW method for $0<\xi<1$ is in between and thus is an improvement on the existing analytical methods. Moreover, we clearly establish the validity of the different RWA methods and the relation of the CHRW treatment to the other RWA methods. Since the CHRW approach is mathematically simple as well as tractable and physically clear, it may be extended to some complicated problems where we cannot do a numerical study. All in all, the approach is not an upgrade patch for the conventional RWA but a much improved innovative RWA. It provides a direct tool for studying the properties of driven tunneling systems within a wide range of parameter. The approach introduced here will prove useful in treating the properties of certain complicated models, in particular in the context of multi-mode driven tunneling quantum dynamics. The work is currently under investigation and will appear elsewhere.

ACKNOWLEDGMENTS

Z.G. L\"u is grateful to C.S. Chu, Y. Teranishi, and R. Chandrashekar for discussions. The work was supported by the MOE ATU Program, Taiwan National Science Council, the NNSF of China (Grants No.11174198 and No.10904091), and the National Basic Research Program of China (Grant No.2011CB922202). Z.G. L\"u gratefully acknowledges J.J. Lin and C.S. Chu for support in all aspects, also acknowledges support from the National Science Foundation
to do research in ICTP (SMR2348 and 2350).

%
%
%
%


\newpage

\begin{center}
{\Large \bf Figure Captions }
\end{center}

\vskip 0.5cm

\baselineskip 20pt

{\bf Fig.1}~~~$P_{\mathrm{up}}(t)$ as a function of dimensionless time $\Delta t$
for $A/\omega= 1, 2$, and $A/\omega=2.5$ in the on-resonance case,
which is shown in Fig. 1(a), Fig. 1(b) and Fig. 1(c), respectively. In each figure,
the result of numerically exact method is plotted by the red dot, those result of the RWA-RF method by the blue dashed dotted line, those of the RWA method in the green dashed line, those of CHRW method by the black solid line.

\vskip 0.5cm

{\bf Fig.2}~~~$P_{\mathrm{up}}(t)$ as a function of dimensionless time $\Delta t$
for $A/\omega= 1$, and $A/\omega=2$ in the near-resonance case($\omega / \Delta=1.2$),
which is shown in Fig. 2(a) and Fig. 2(b), respectively.  In each figure,
the result of numerically exact method is plotted by the red dot, those result of the RWA-RF method by the blue dashed dotted line, those of the RWA method in the green dashed line, those of CHRW method by the black solid line.

\vskip 0.5cm

{\bf Fig.3}~~~$P_{\mathrm{up}}(t)$ as a function of dimensionless time $\Delta t$
for $A/\omega= 1 $, and $A/\omega=2$ in the near-resonance case($\omega/ \Delta=0.8$),
which is shown in Fig. 3(a) and Fig. 3(b), respectively.In each figure,
the result of numerically exact method is plotted by the red dot, those result of the RWA-RF method by the blue dashed dotted line, those of the RWA method in the green dashed line, those of CHRW method by the black solid line.

\vskip 0.5cm

{\bf Fig.4}~~~$P_{\mathrm{up}}(t)$ as a function of dimensionless time $\Delta t$
for $A/\omega= 1, 2, 2.5$, and $A/\omega=6$ in the far off-resonance case($\omega/ \Delta=5$),
which is shown in Fig. 4(a),Fig. 4(b), Fig. 4(c) and Fig. 4(d), respectively. The dynamics of the other far off-resonance case $6\omega=\Delta$ is shown in Fig.4(e)($A/\omega=2$) and Fig4.(f)($A/\omega=6$).  In each figure,
the result of numerically exact method is plotted by the red dot, those result of the RWA-RF method by the blue dashed dotted line, those of the RWA method in the green dashed line, those of CHRW method by the black solid line. Inset shows the detailed comparison of these methods in the aspect of the structure of short-time dynamics.

\vskip 0.5cm

{\bf Fig.5}~~~$P_{\mathrm{up}}(t)$ as a function of dimensionless time $\Delta t$ with $A/\omega=2.4048$($J_0(2.4048)\simeq 0$) for different driving frequencies, $\omega/\Delta=2$ (a), $\omega/\Delta=6$ (b), $\omega/\Delta=10$ (c), and $\omega/\Delta=20$ (d). The red dot gives the predicted exact dynamics from the numerical method, the green dashed line gives the predicted dynamics from the RWA method, the blue dashed dotted line gives the predicted dynamics by the RWA-RF method, and the black solid line shows the CHRW result. Because of $J_0(2.4048)\simeq 0$, the RWA-RF method shows the CDT for any driving frequency, which is not accurate and needs improvement. Inset shows the whole comparison of these methods.

\vskip 0.5cm

{\bf Fig.6}~~~Regions of validity for different RWAs for zero static bias case. The CHRW method is an improved RWA method in this work. The Rabi-RWA stands for the traditional RWA method, not RWA-RF (the results of the RWA-RF method by Eq. \ref{rrwa}).  The RWA method give good results in near-resonance or on-resonance case for weak driving limit(in comparison with numerical exact results,  the deviation of amplitudes and oscillation frequencies given the RWA method is limited by $5\%$) \cite{foot}. The axes are the amplitude $A$ of the driving field and the tunneling strength $\Delta$ of the TLS, both normalized to the frequency $\omega$ of the driving field. The Rabi-RWA region is shown in dense shaded area. The CHRW region is described by the sparse shaded area. The red rectangular area indicates the parameter region with the phenomena of much large suppressed tunneling amplitude. Meanwhile, we have shown all calculated data points in this figure by different symbols (the data of Fig.1 by the solid cycle; the data of Fig. 2 and Fig. 3 by the solid triangle; the data of Fig.4 by the solid star).

\newpage

\begin{center}
{\Large \bf Table }
\end{center}

\begin{table}[htbp]
\caption{Summery of the results of Fig. 5 ($A/\omega=2.4048$)} 
\centering 
\begin{tabular}{c|c|c|c|c|c|}
\hline\hline 
Figure & $\frac{\omega}{\Delta}$ & RWA  & RWA-RF & CHRW & Exact \\ [0.5ex]
\hline
Fig.5(a) & 2&  NO CDT & CDT $P(t)=0$ & NO CDT & NO CDT\\
Fig.5(b) & 6&  NO CDT & CDT $P(t)=0$ & NO CDT & NO CDT \\
Fig.5(c) & 10& NO CDT & CDT $P(t)=0$ & CDT*& CDT* \\
Fig.5(d) & 20& NO CDT & CDT $P(t)=0$ & CDT*& CDT* \\[1ex]
\hline
\end{tabular} \\
Note: The CDT* means an oscillation with a nonvanishing small amplitude. The CDT of the RWA-RF method is $P(t)=0$.
\label{tab:hresult}
\end{table}


\begin{thebibliography}{99}
\bibitem{foot} Note that in the off-resonance case $|1- \Delta/\omega|>0.2$, the Rabi-RWA dynamics exhibits large difference from numerical exact one. Thus, in the Fig. 6, we show a small dense shaded area for Rabi-RWA with good performace. While the amplitude of the RWA-RF always keeps one, we do not show its valid area in the figure and its prediction of the CDT for arbitary $\Delta/\omega$ is false as long as $J_0(A/\omega)=0$.
\end{thebibliography}
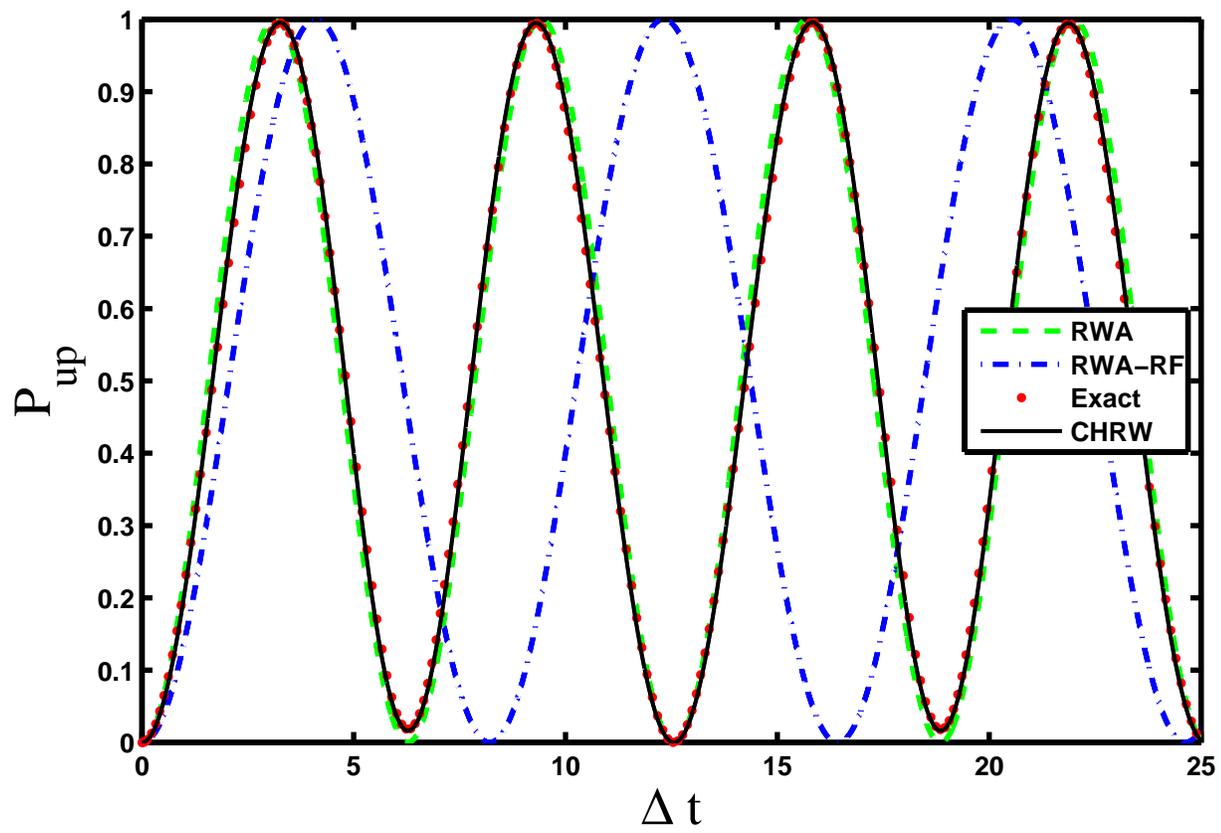
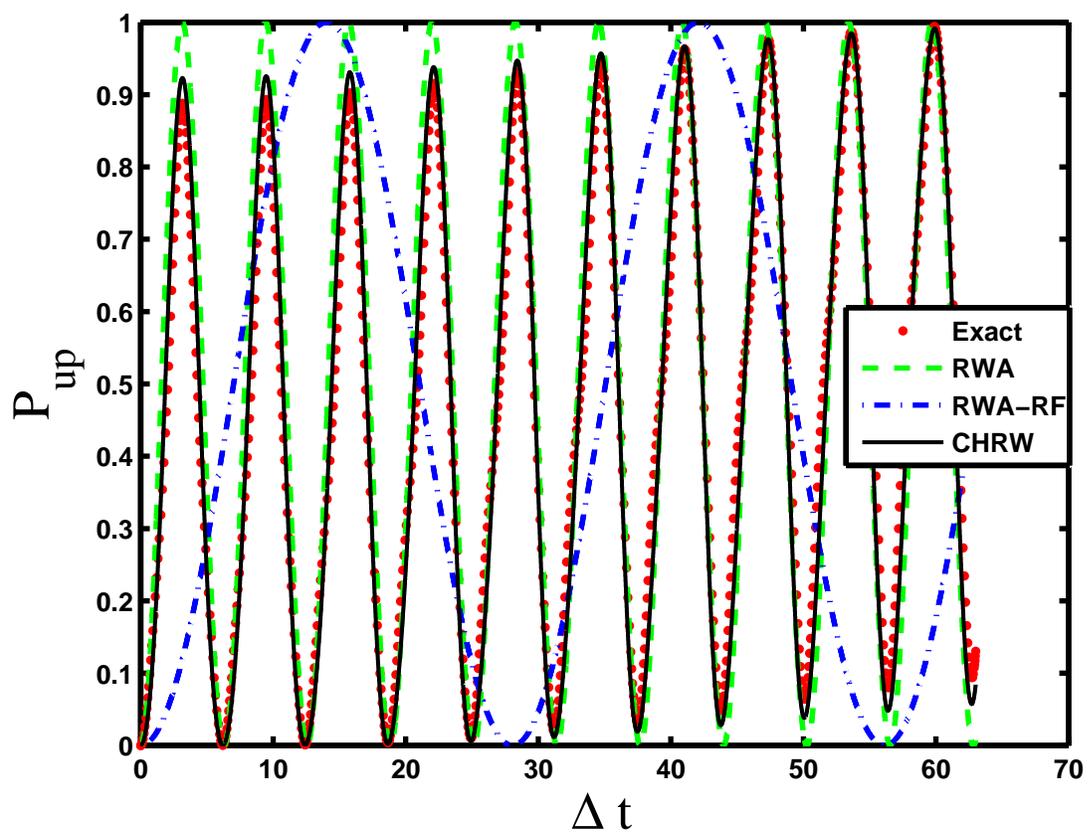
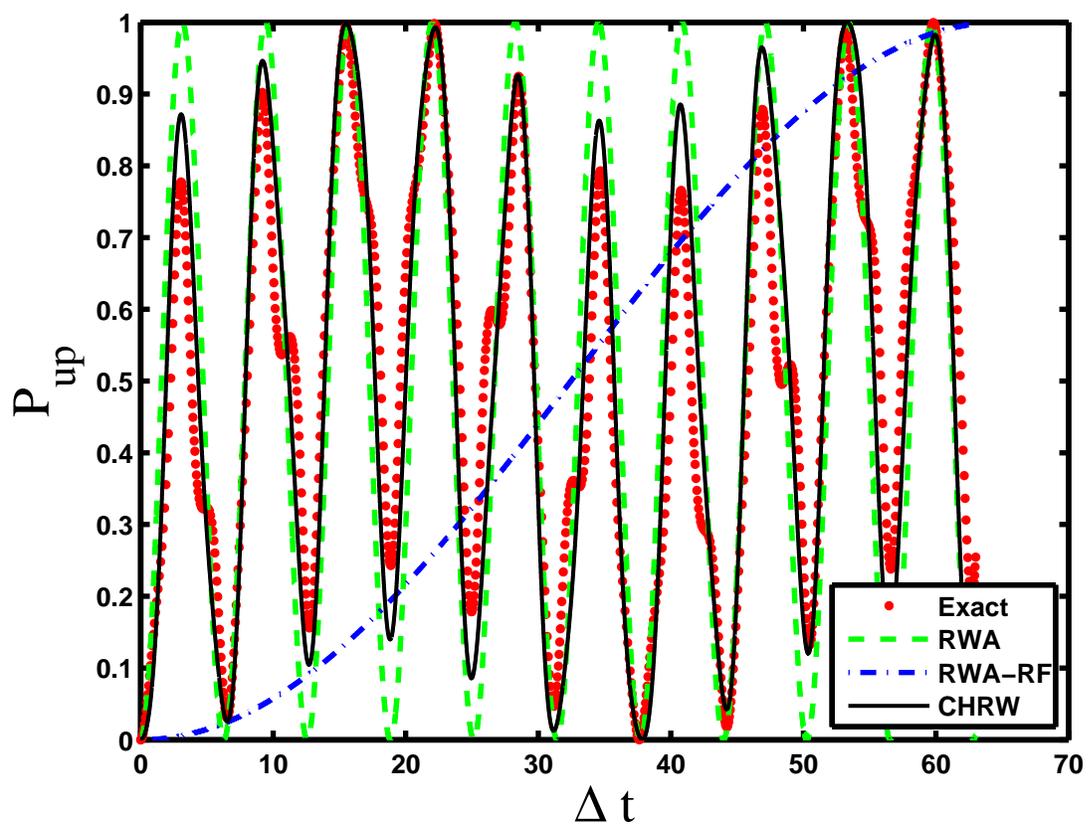
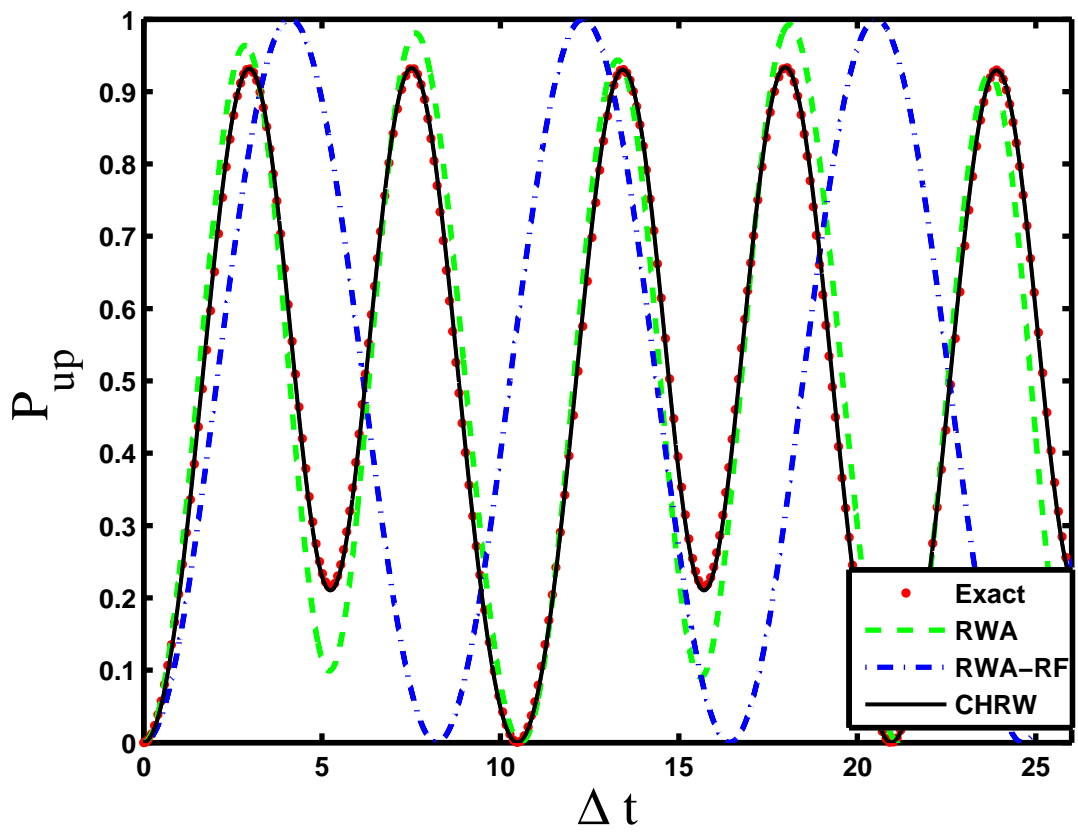
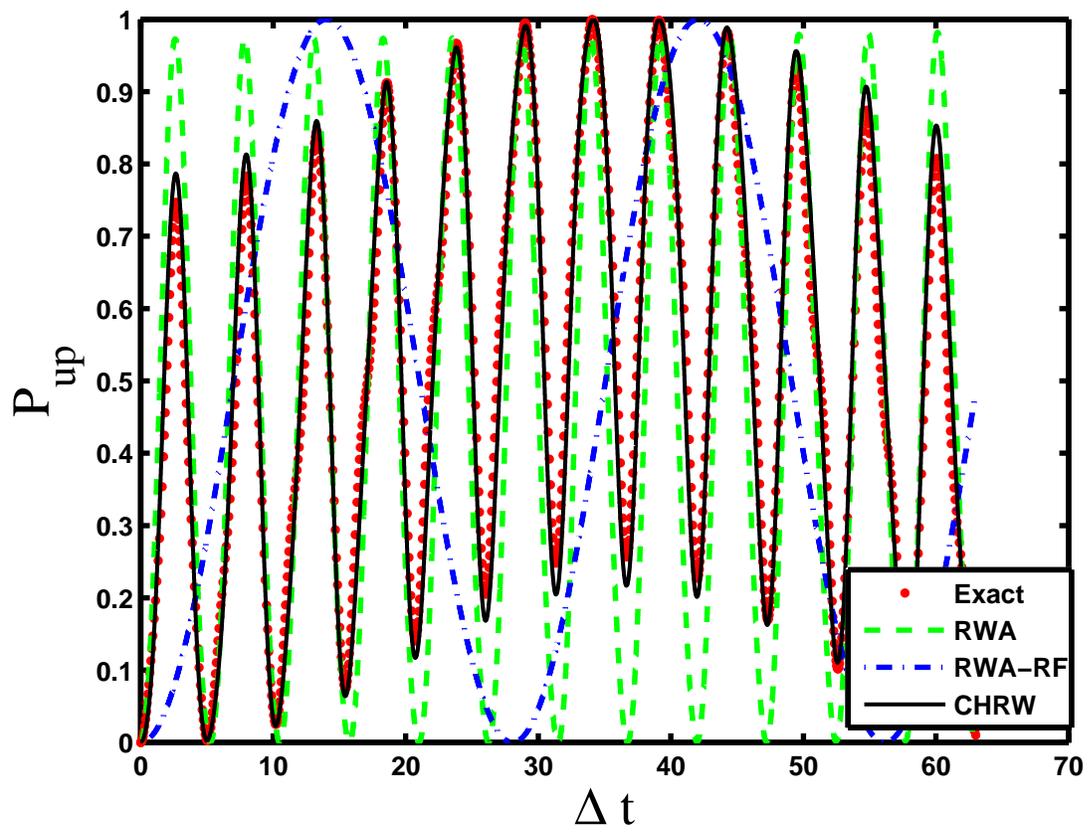
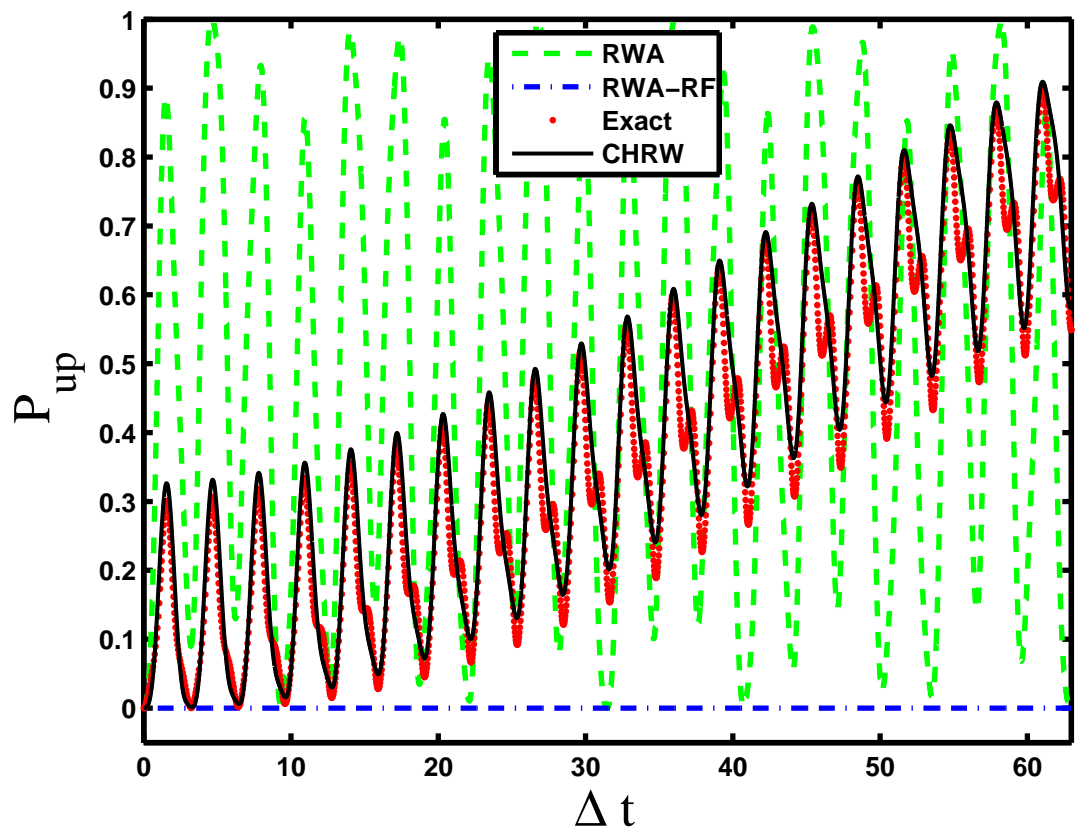
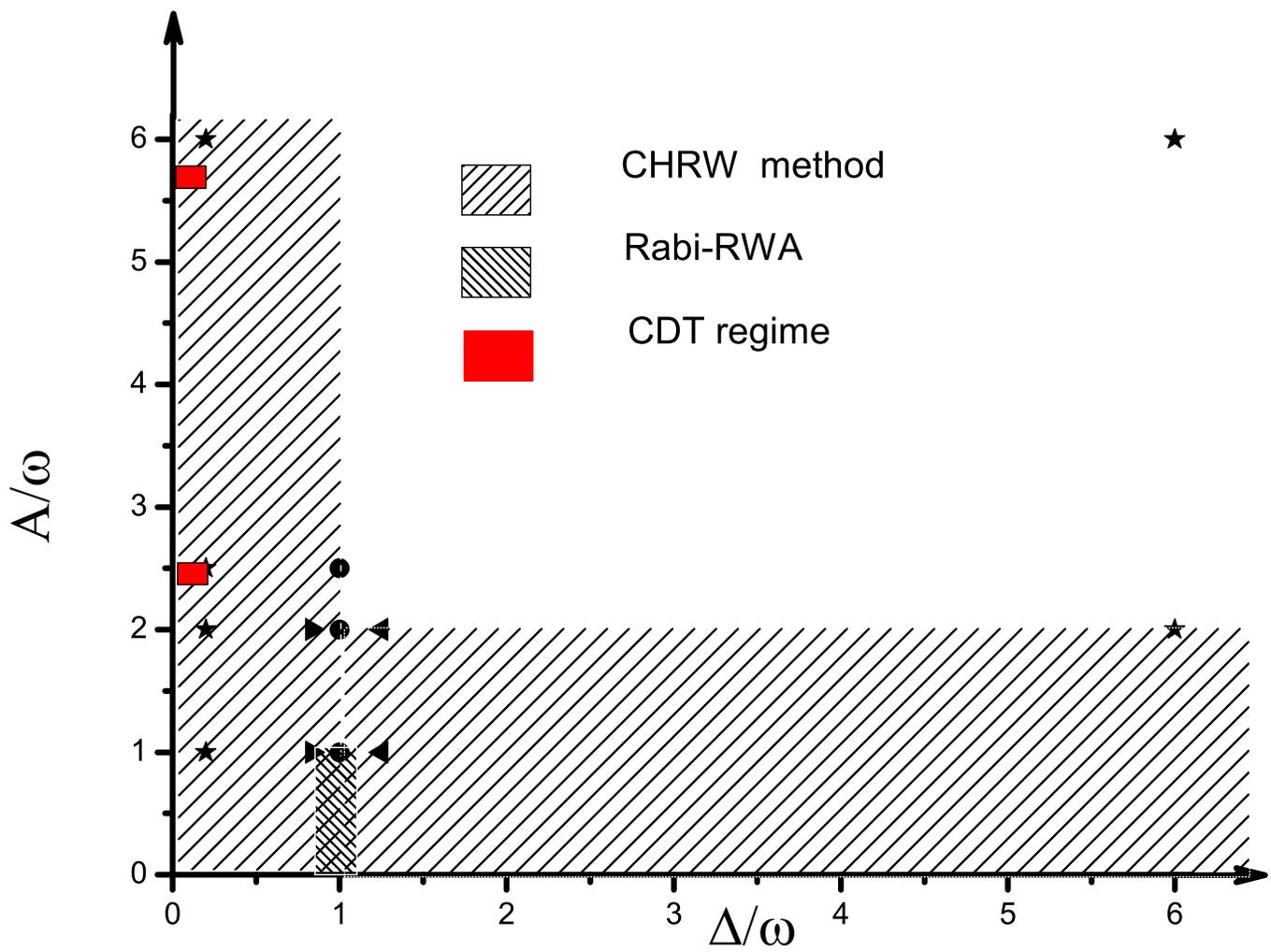
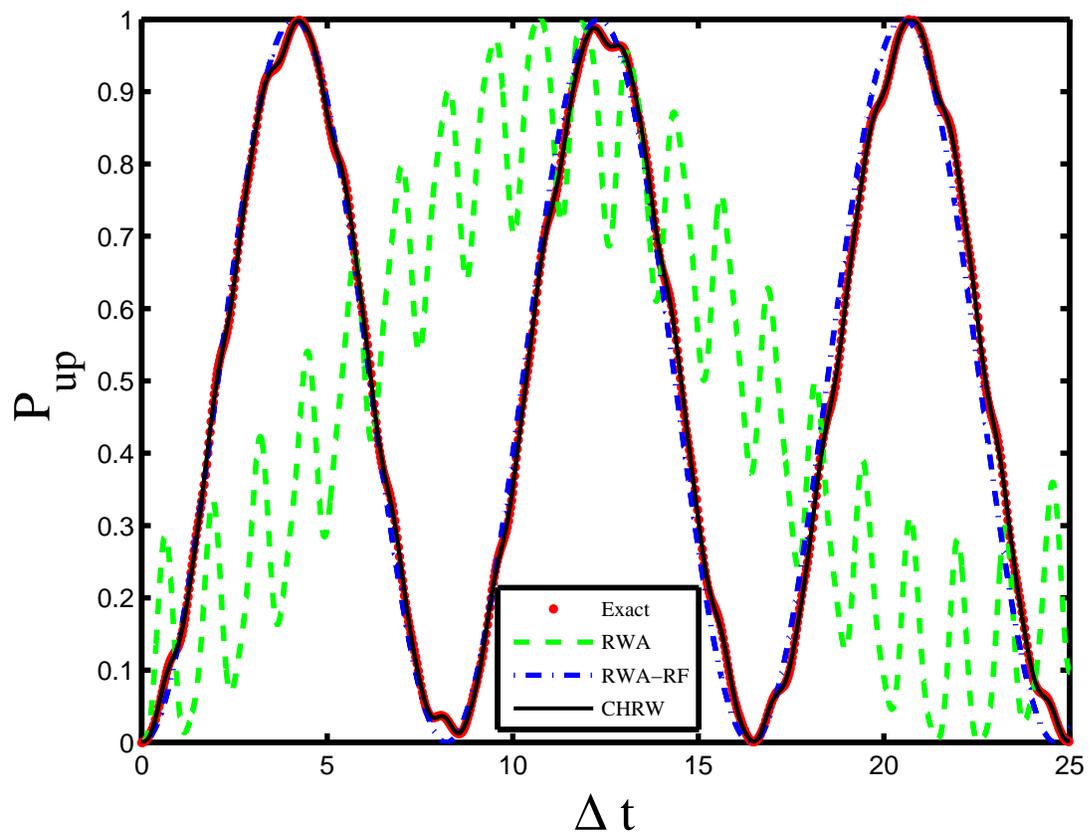
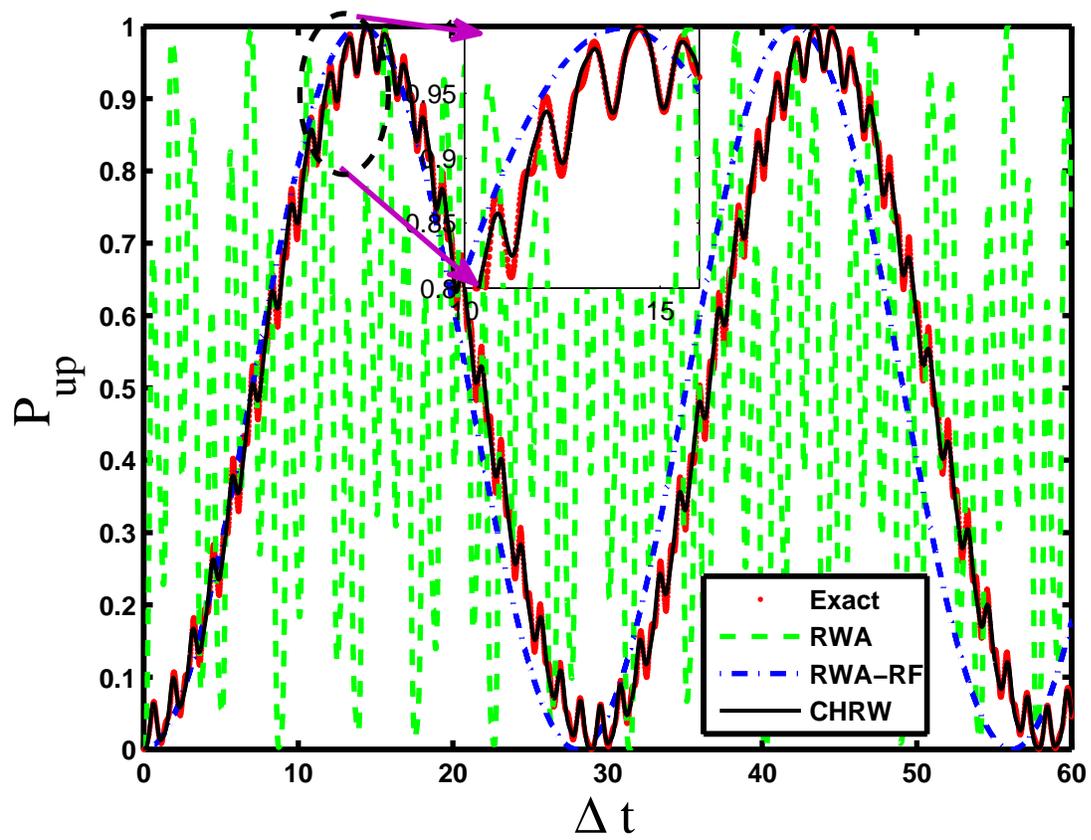
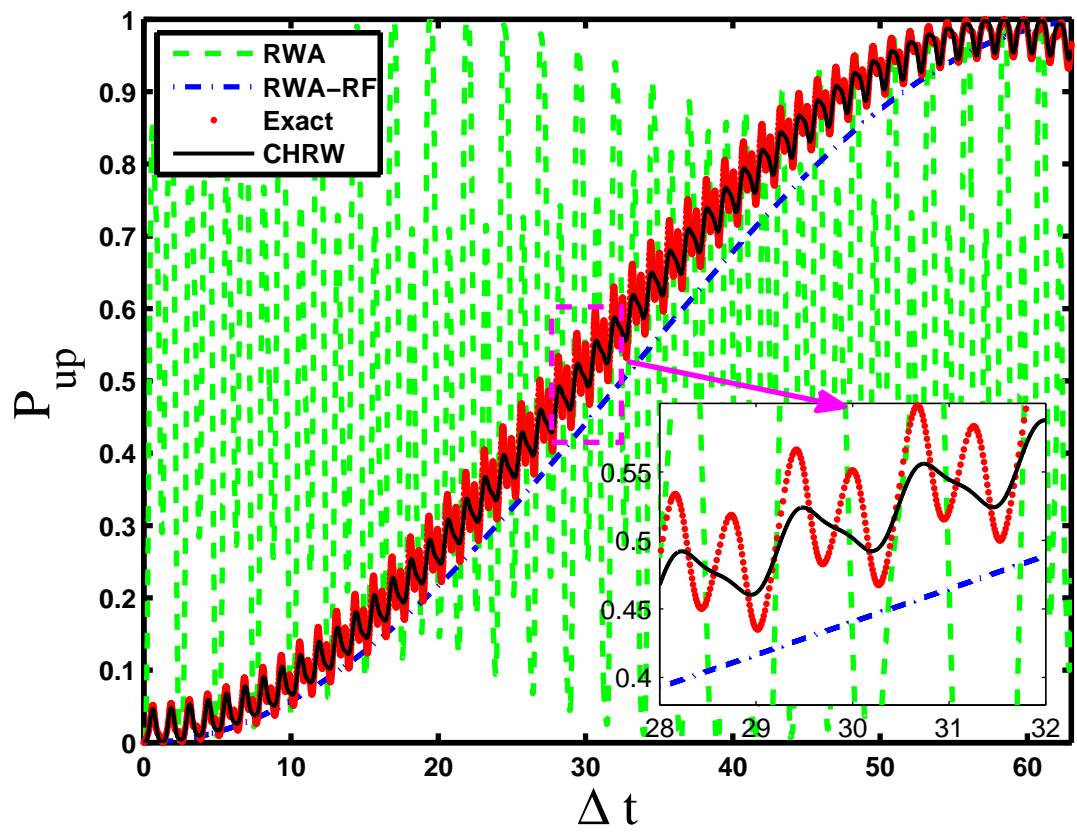
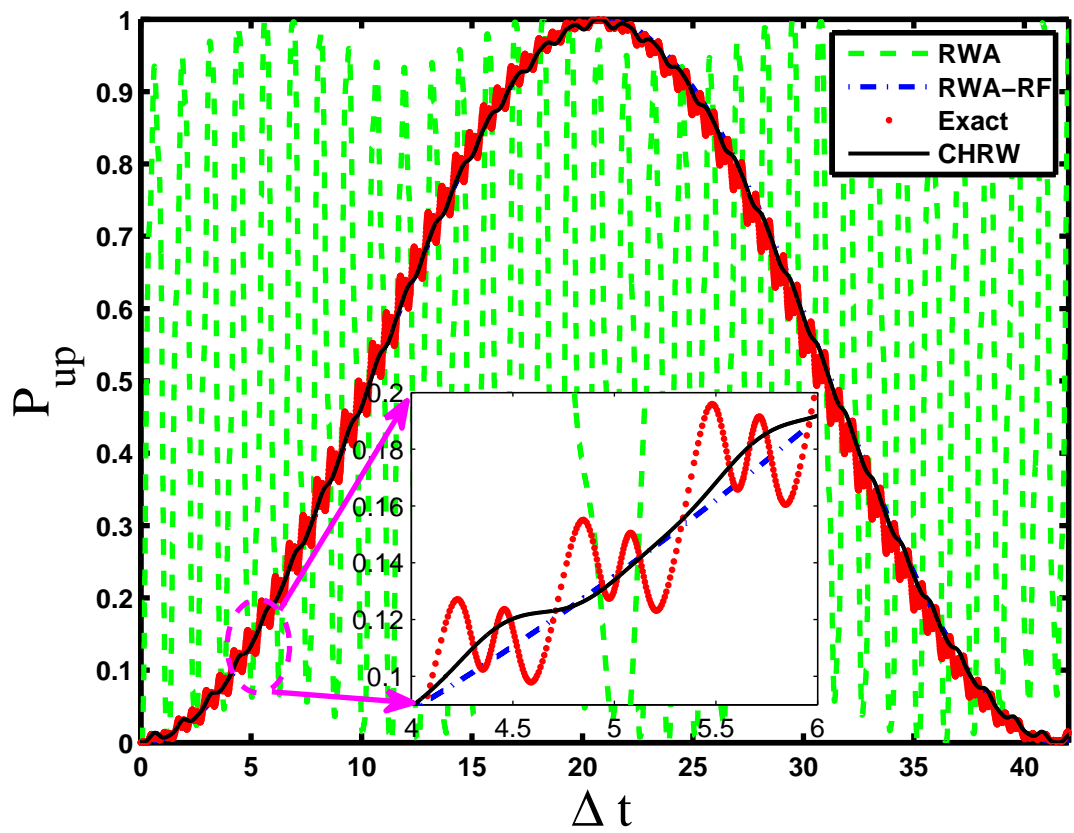
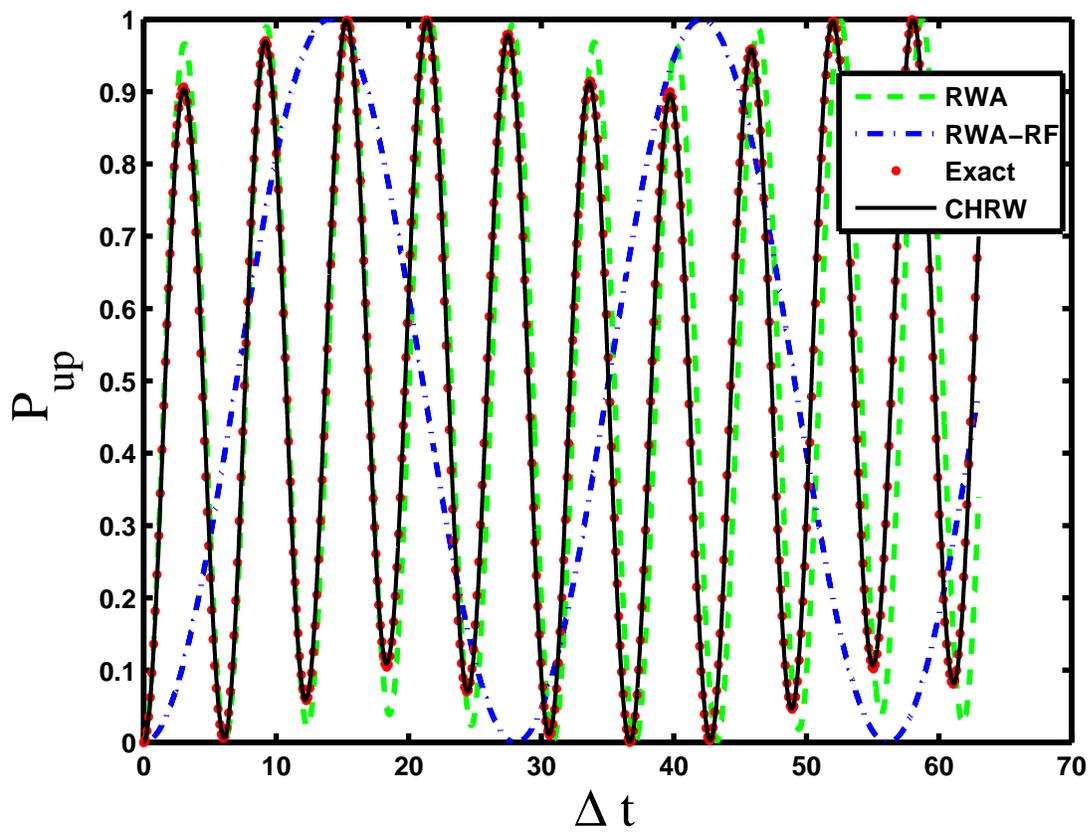
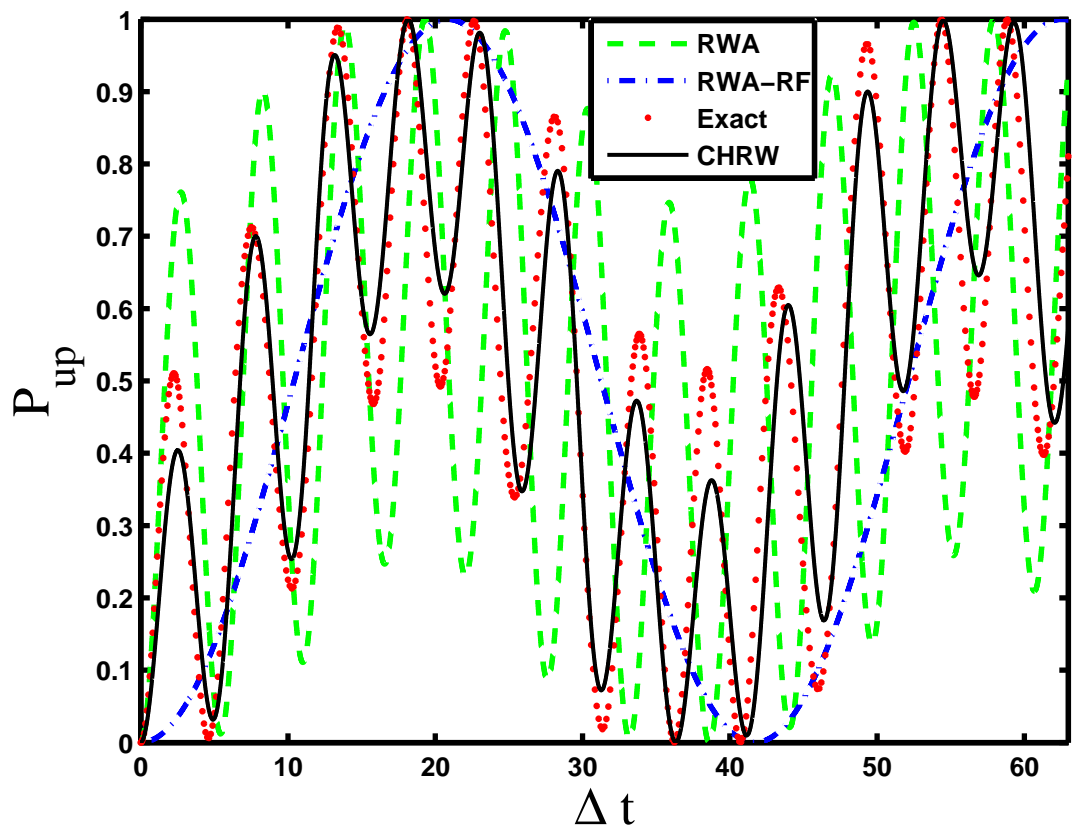
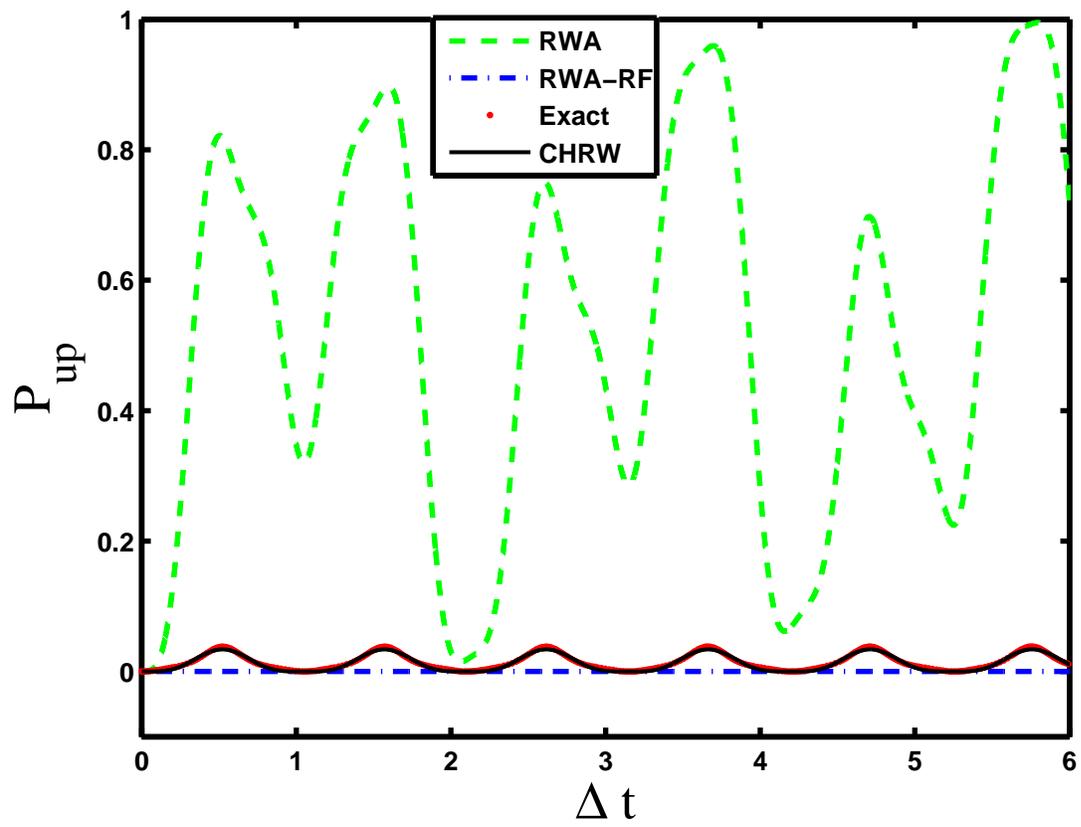
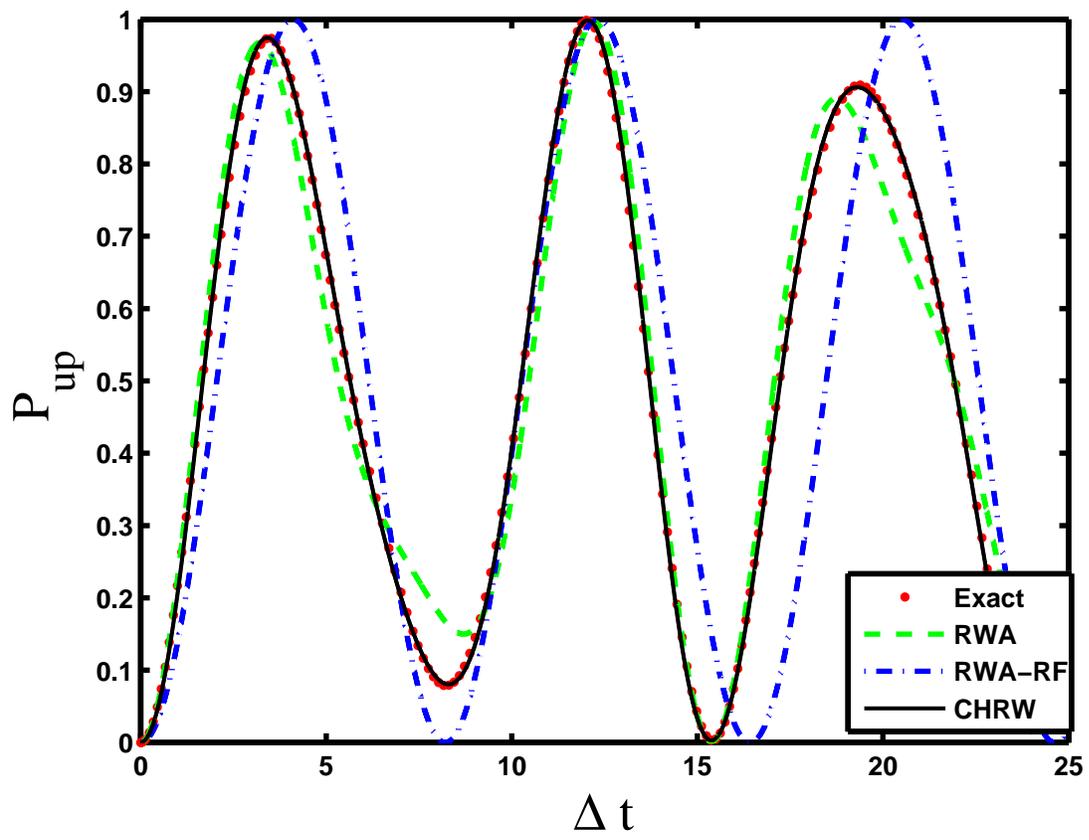
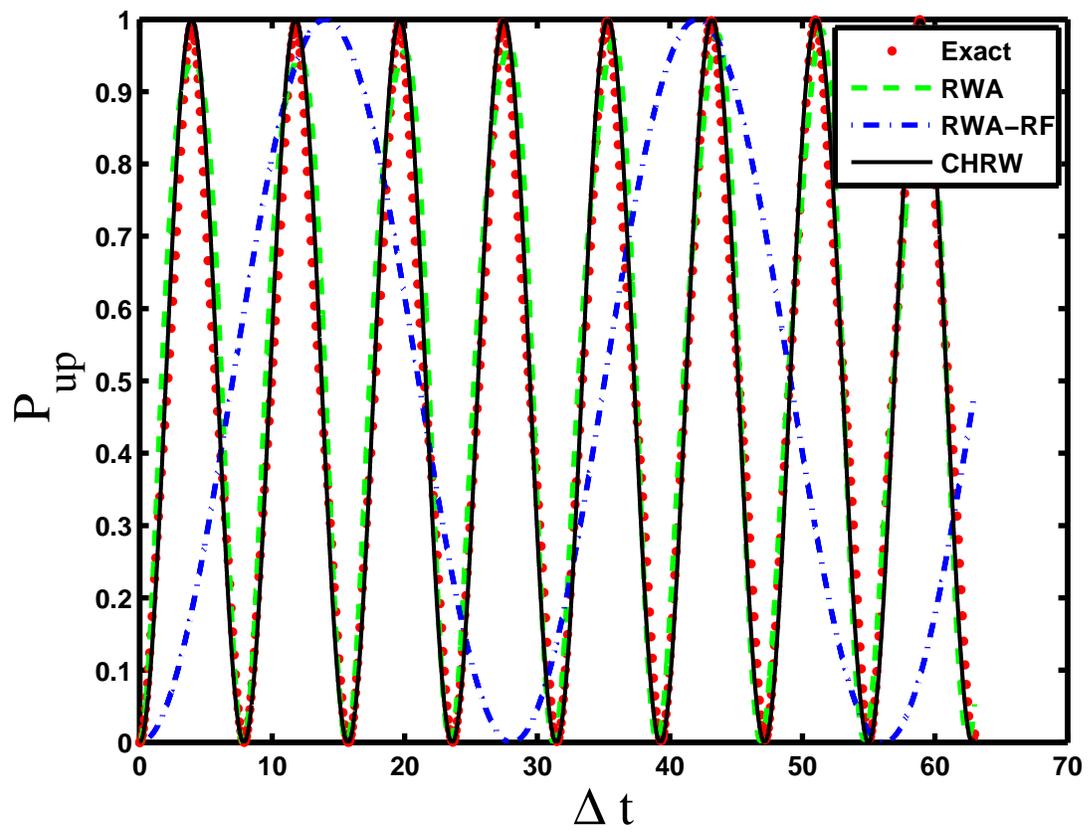
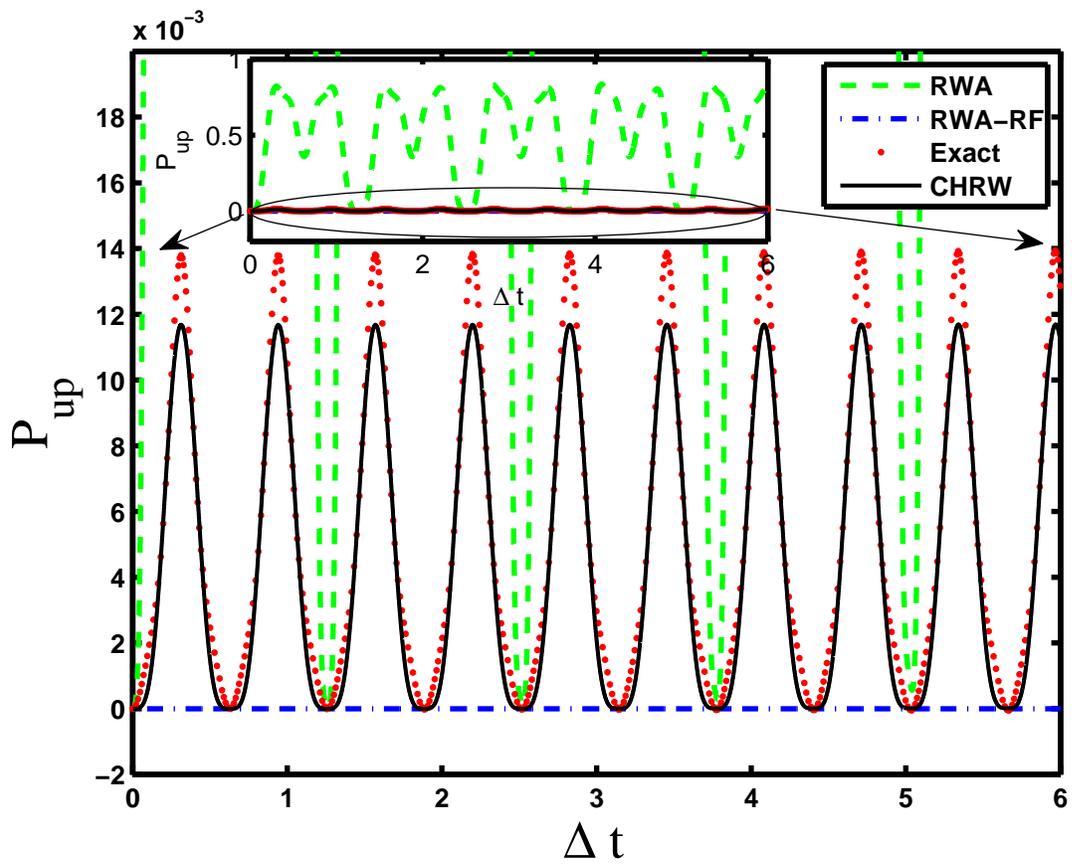
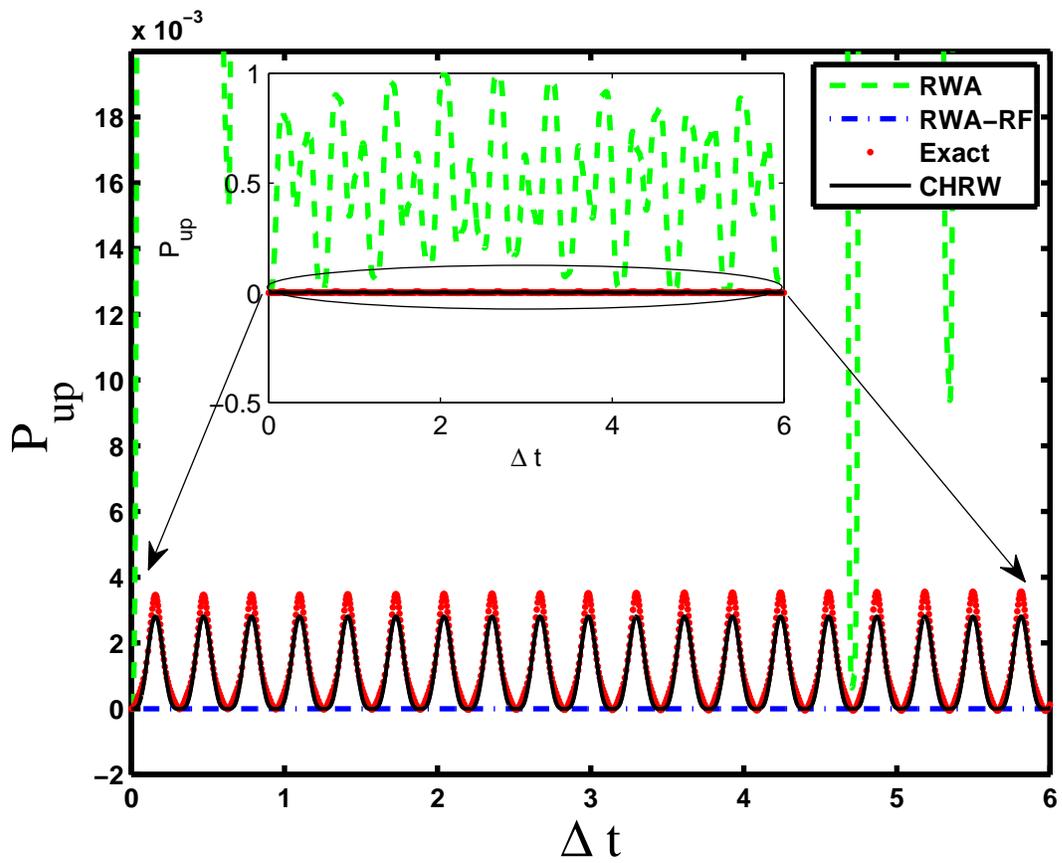
\end{document}